# Metal-Organic Frameworks in Semiconductor Devices: Recent Advancements and a Bright Future

Ranjeev Kumar Parashar, Priyajit Jash, and Prakash Chandra Mondal*

Department of Chemistry, Indian Institute of Technology, Kanpur, Kanpur-208016, India

E-mail: pcmondal@iitk.ac.in (PCM)

## Abstract

Metal-organic frameworks (MOFs) symbolize the particular class of hybrid crystalline, nano-porous materials made of either discrete metal ions or clusters with organic linkers. Past studies on MOFs-based materials largely focused on porosity, chemical and structural diversity, gas sorption, sensing, drug delivery, catalysis, and separation applications. However, initial efforts either neglected or have not gained much attention to refine the electrical conductivity of MOFs materials. MOFs reported earlier with poor electrical conductivity ($\sigma < 10^{-7} – 10^{-10}$ S cm$^{-1}$) impeded to employ in electronics, optoelectronics, and renewable energy storage applications. To overcome this issue, the MOFs community has been engaged in improving electrical conductivity by adopting several intriguing strategies. We shed light on the charge transport mechanisms which are mainly two processes, either through a bond or through space. This review aims to showcase the current scenario on creatively designed MOF materials followed by fabrication advancement of high-quality molecular thin films, and semiconductor device fabrication for stimuli-responsive current-voltage (I-V) studies. Overall, the review addresses the pros and cons of the MOFs-based electronics, followed by our prediction on improvement MOFs composition, mechanically stable interfaces, device stacking, further relevant experiments which can be of great interest to the MOFs researchers in improving further devices performances.

**Keywords:** electronics devices; charge-transport; electrical conductivity; electrode/MOFs/electrode junctions, current-voltage characteristics

# 1. Introduction

*Metal-Organic Frameworks* (MOFs) are suitably designated as 'Holy Grail', as these versatile materials have revealed historical evolution in material science over recent decades. MOFs are composed of multitopic ligands called 'arms' and metal ions or metal-ions based clusters called 'nodes'. Crystalline bulk MOFs are commonly synthesized via solvothermal method performed at high temperature followed by slow cooling (2-5°C/h). During this process, 'nodes' and 'arms' coordinate each other via a self-assembly process leading to the formation of highly ordered, crystalline cage-like MOFs. The journey of MOFs domain begins in 1995, by Prof. Yaghi and his research team after the discovery of the first MOFs using 4,4'-bpy linkers as 'arms' and $Cu^{2+}$ as 'nodes' held together by a coordination bond [1,2]. Since then, MOFs have drawn enormous attention to the researchers for the diverse areas of scientific applicability. Undoubtedly, beyond synthetic deliberation, the crystal engineering of MOFs ignites the choice of different organic ligands with suitable functional groups and the corresponding metal ions with distinct directionality, geometry, and functionality [3]. Hence, the infinite possibilities for MOF designs can be flagged for desired technological importance. Indeed, the exploitation of coordination bonds leads to forming MOFs of 1, 2, and 3-dimensional networks (1D, 2D, 3D) through the assemble of molecular building blocks toward different supramolecular architectures. Among them, 3D MOFs are known as promising materials for superior physical properties, including a longer pore diameter which is good enough for small molecules storage, gas adsorption, sensing, catalytic transformations, photo-switching, bio-medical applications [4–8]. Despite the easy preparation, low-cost production, high crystallinity that MOFs offer, their poor electrical conductivity that arises due to weaker d-p orbitals interaction between transition metal ions and organic linkers make them inadequate for microelectronic applications. Transitional metal ions-MOFs are found poor conductors over the lanthanides-MOFs. We demonstrate design strategies of MOFs-based molecular electronics followed by direct current (DC) based measurements to understand the current flowing in the circuit response with respect to the applied bias. In this review, we intend to correlate the fabrication process of bulk MOFs into surface-mounted metal-organic frameworks (SURMOFs) towards highly conducting thin film-based device integration. Additionally, we have emphasized the paradigm transformation of basic science to advanced electronic applications. The schematic overview of several depositions and molecular assembly techniques on solid substrates and the roadmap of MOF for the application of molecular electronics is shown in **Figure 1**.

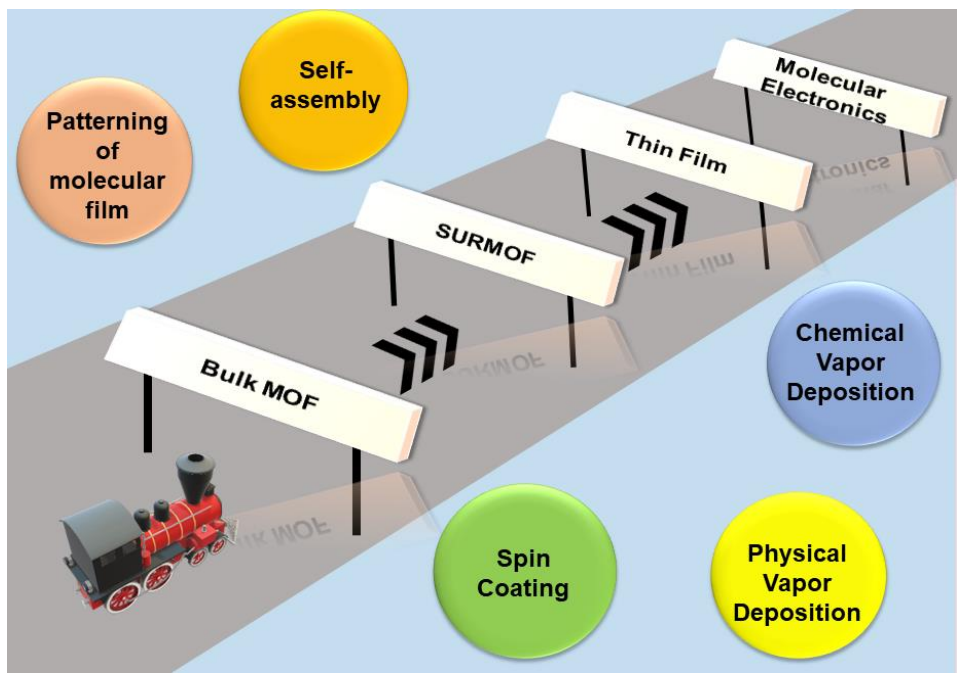

**Figure 1**: Schematic representation of multiple routes for assembling functional molecules and deposition on solid substrates envisioned for MOF fabrication towards molecular electronics application.

## 2. MOFs/Electrode Interfaces Design strategies: SURMOFs vs. Bulk MOFs

Successful design, synthesis, and engineering of solution-based MOFs chemistry have propelled a smart research domain, called SURMOFs, led by the team of Prof. Christof Wöll and his team. The team has established a methodology to manipulate surface chemistry for creating unique SURMOFs [9–11]. SURMOFs are thin, crystalline, and have a high degree of structural ordered MOF film having adjustable defect density [12,13]. SURMOFs growth is likely to occur either by layer-by-layer (LbL) technique like spin-coating or drop-casting method onto different substrates. MOFs or coordination polymers (CPs) building blocks can be deposited onto a pre-functionalized substrate of template layer coupled to the stepwise LbL method [14–18]. There have been various successful deposition methods that can attach MOFs onto to pre-functionalized substrates (**Figure 2**). Choosing an apposite substrate and smart surface modification is imperative for the MOF thin films deposition [10]. So far, several strategies have been developed over the choice of the substrate like a coating of planar solid (Si, Au, FTO, ITO, etc.), flexible (plastic), and nonplanar substrates such as metal oxide or polymer particles, metal foams, etc.[19–21]. In another way, the modification of the substrates is simultaneously important, in specific for the quasi-epitaxial deposition of thin films. Towards that goal, many synthetic approaches, including liquid-phase epitaxy (LPE), chemical vapor deposition (CVD), atomic layer deposition (ALD), bottom-up, substrate-seeded heteroepitaxy (SSH), and electrochemical fabrication (ECF), have been established for the fabrication of MOF thin films [21,22]. Recently, LPE was found to be one of the smart synthetic strategies for SURMOFs as it can enable the formation of hetero multilayers as well as provides well-defined electrical contacts which can influence the device performances [19,23].

During the LPE process for SURMOFs, the reactants are separated individually compared to the ''one-pot'' reaction scheme applied for the solvothermal synthesis of powder MOF [10]. As one of the well-known MOF, the HKUST-1 was made as a highly oriented crystalline film with a changeable thickness on a carboxylic terminated self-assembled monolayers (SAMs) on the Au substrate using stepwise LPE technique by the integration of copper nodes and trimesic acid [24–26]. The SAMs are considered ideal and surface model systems for many theoretical and experimental studies including nanotechnological applications [27–30]. Several research groups have reported the fabrication of SAMs-based molecular electronic and spintronic devices incorporating molecules as circuit elements [29,31–34]. The obtained SURMOF of different thicknesses was tested in Hg-based tunneling junctions as a bottom electrode to investigate the hopping charge transport mechanism (vide infra).

LPE method was further optimized for the fabrication of ultrathin films (thickness < 30 nm) containing a series of $Fe^{II}$ Hofmann-like coordination polymers (CPs) having the composition of metal-cyanide-metal layers with the formula $[Fe(L)_2\{Pt-(CN)_4\}]$, where L is the nitrogen-based aromatic pillaring ligands coordinated to octahedral $Fe^{II}$ metal ions (**Figure 3**) [35]. The series (L = pyridine, pyrimidine, and isoquinoline) correlate the consequence of the axial nitrogenated ligand with adjustments to their structural response to the guests or to mitigate the electrical resistance. However, the fabrication step is still riddled with the rate of the growth i,e. the kinetic problem, which is another field of extensive study. Tuning of functional groups like terminal components of the SAMs can initiate the nucleation, which presumably makes easy coupling of either the metal or metal-oxo nodes and organic linkers for the substantial growth of the MOF thin films [36]. As per the appealing features, the fabrication step must have to overcome each individual challenge substantially. The electrochemical deposition (ECD) technique is another simple and robust tactic for the design of MOF thin films. It has been categorized into three methods: 1. Anodic deposition, 2. Cathodic deposition and, 3. Electrophoretic deposition. The main advantage of using the ECD method is due to no capping reagent or surfactant or any other dispersion agent is involved [37]. The chemical vapor deposition (CVD) technique can also be used for the same purpose. Since it is based on adsorption and subsequent chemical reaction of vapor with the substrate surface to harvest with strongly controlled coating dimensions [38]. Another decisive feature of thin-film MOF is the introduction of defects densities in a controlled way during film formation. For instance, in the case of SURMOF, using defective linkers, low defect densities can be produced in a straightforward approach to unravel the daunting task between defects and materials [39]. Recently, Heinke and co-workers made semiconducting structurally well-defined, heterojunctions using LbL deposition of MOF thin films[40]. X-ray diffraction study revealed perfect epitaxy through changing the lattice constants of the two different MOFs. Deposition of top electrodes permits the construction of p–n and n–p devices which showcase a high $I_{on/off}$ ratio up to 6 orders of magnitude.

Therefore, it has become the current thrust to produce high-quality thin-film MOF consisting of enhanced electrical conductivity, charge carrier mobility, charge separation efficiency, and interface stability, etc. Although that is not the initial scope of this review, we intend to showcase the standard shift of the chemical features from bulk to the advanced thin-film MOFs on how the efficacy of electrical charge transport can be improved. In this regard, Dincă and coworkers have synthetically developed binary alloy-based MOFs using 2,3,6,7,10,11-Hexaiminotriphenylene (HITP) as an organic linker and CuNi, CoNi, CoCu, respectively as the inorganic precursor which engenders electrical conductivity of beyond four orders of magnitude [41]. They could modify the electrical conductivity and the bandgap of the MOFs by varying the metal ions, which makes alterations in the interlayer spacing fortifying the bulk MOF structure-function relations. Recently, Rambabu *et al.* prepared electrical conducting Li-ion intercalated MOF that showed electrode behavior ($Li_x$-M-DOBDC where, x = 1,1.5,2; M=$Mg^{2+}$or $Mn^{2+}$; $DOBDC^{4-}$ = 2,5-dioxido-1,4-benzenedicarboxylate, that exhibits a million-fold increment of electrical conductivity ($10^{-7}$ S/cm) for $Li_2$-Mn-DOBDC as compared to reticular analogues ($10^{-13}$ S/cm) like $H_2$-Mn-DOBDC and $Li_2$-Mg-DOBDC (**Figure 4**) [42]. They have speculated that the massive rise of conductivity is aroused by the effective self-exchange electronic transition through the redox mediator, making efficient electrode material for capacitive charge storage.

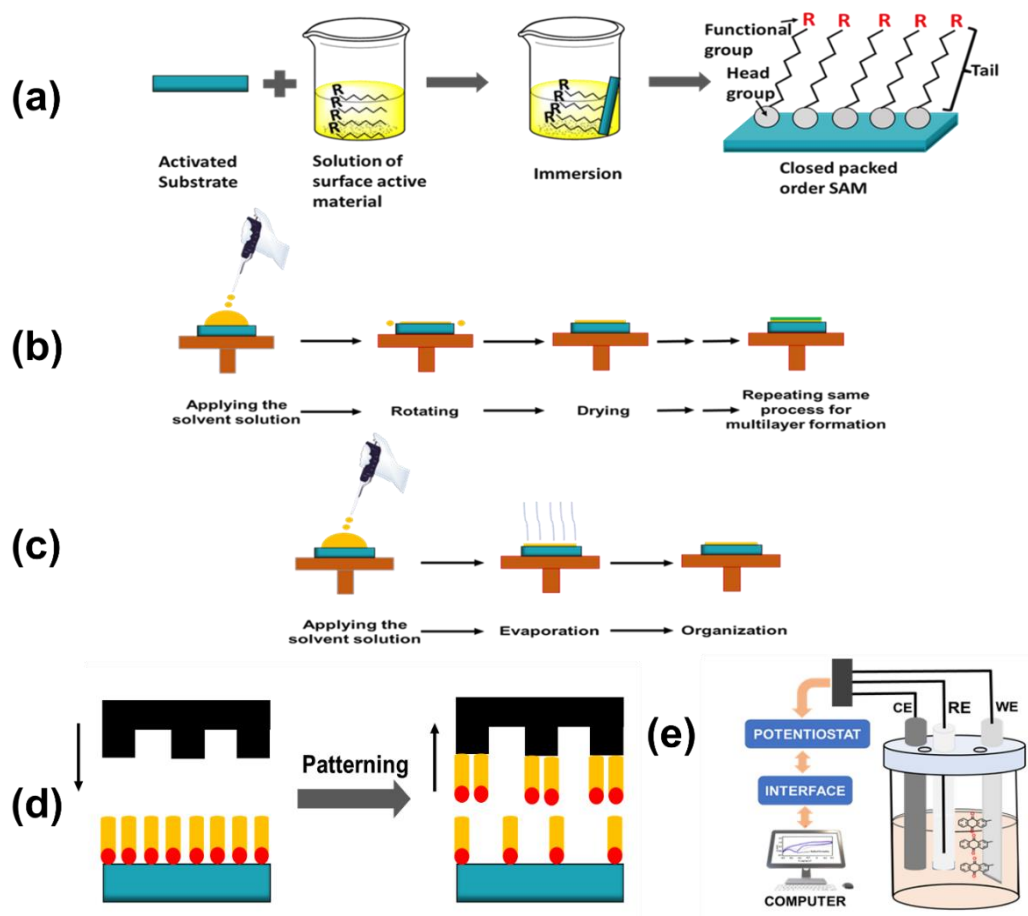

**Figure 2.** Schematic illustration describing the varied chemical and physical routes for fabricating surface-confined MOFs on pre-functionalized substrates. The techniques include (a) self-assembly, (b) spin coating, (c)

drop-casting, (d) photo-patterning, and (e) potential-driven electrochemical deposition on working electrodes in a three-electrode electrochemical cell. Vapor depositions are excluded here.

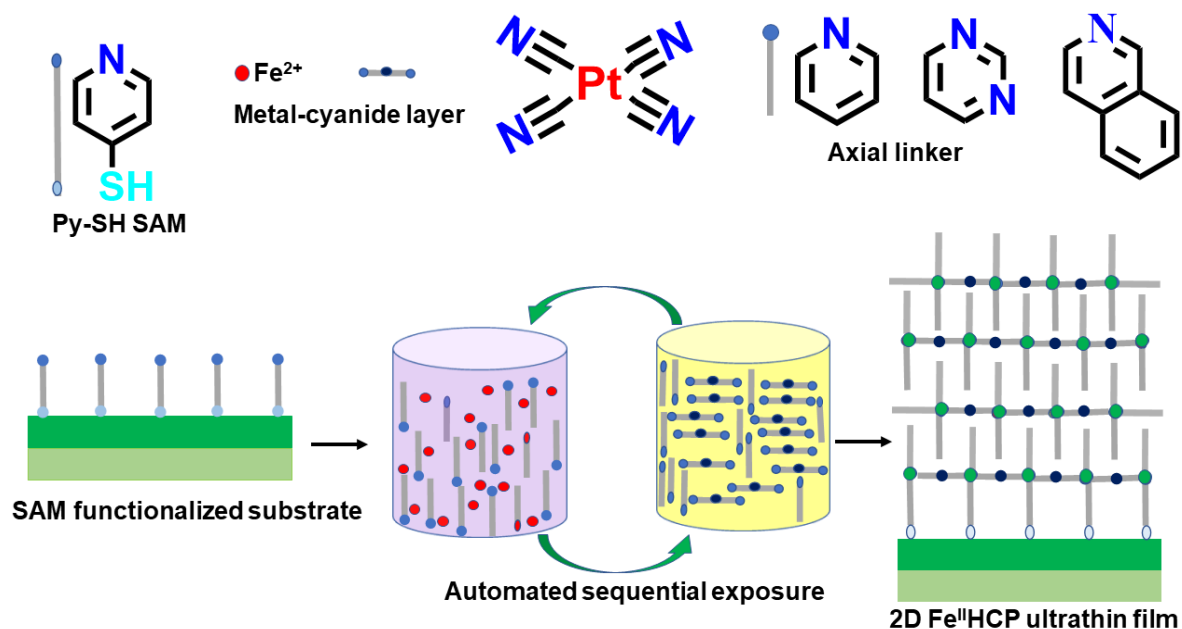

**Figure 3.** Automated LPE ultrathin film synthesis method of [Fe(L)$_2${Pt(CN)$_4$}] using LPE method. The SAM-functionalized on Si/Au substrate is successively absorbed in ethanol solutions of the molecular ingredients like at the beginning, Fe$^{II}$/L, and then the axial ligands [Pt(CN)4]$^{2-}$/L with intermediate washing steps using pure ethanol. Sequential cycling for controllable film thickness has been completed using an automatic dipping system to confirm reproducibility.

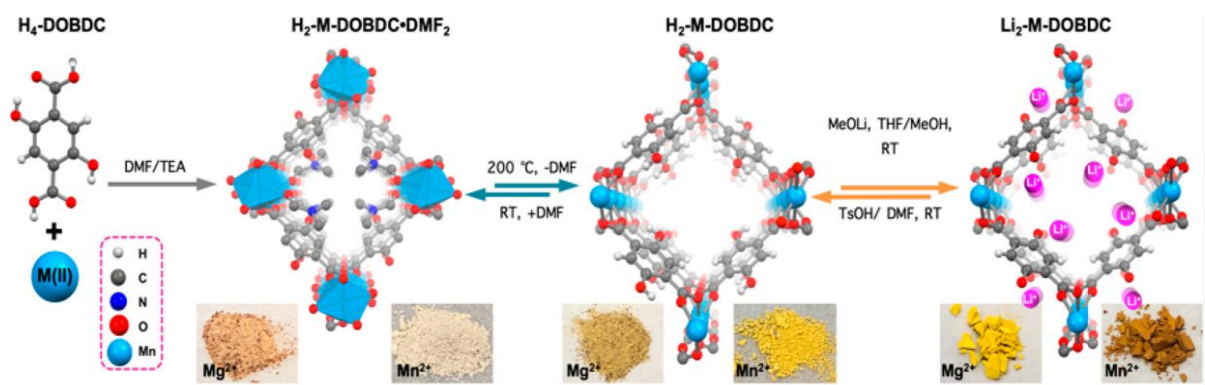

**Figure 4.** Schematic diagram of the fabrication process for H$_2$-M-DOBDC·DMF$_2$, H$_2$-M-DOBDC, and Li$_2$-M-DOBDC (M = Mn, Mg). Small insets are the Optical micrographs presenting the color and morphology of the compound in powder form for individual composition. Reproduced with permission from ref. [42], copyright 2021, American Chemical Society.

# 3. Breaking of bottleneck in electrical conductivity

MOF design and synthesis have advanced dramatically in recent years. Initially, a carboxylate group-containing linker and metal ions were extensively explored to synthesize MOFs with permanent porosity that was best suited for conventional applications[43,44]. Despite the plethora of success of such MOFs in conventional applications, nevertheless, employing them in electronic devices remains a challenge due to their extremely low conductivity of $10^{-10}$ S/cm or even lower. Because majority of such MOFs have a wide band gap (>3 eV), are composed of metal clusters and redox-inert linkers, and their metal and ligand orbitals do not overlap sufficiently to allow for effective charge transfer across the framework. As a result, they are often electrically inactive. Thereby, converting electrically inert MOFs into ones that are electrically conductive is an exceedingly daunting task.

However, the MOF community has conducted significant research work to conquer the MOFs' electrical conductivity challenges. Finally, in 2009, a crucial milestone was accomplished as Long, and Co-workers reported the first electrically conductive MOF. With this finding and the ramification of recent advances, conductive MOFs have emerged as a unique class of electronic materials and a viable alternative to Complementary metal-oxide–semiconductor (CMOS) for fabricating electronic devices. Presently, MOFs have been established as versatile electroactive materials which are frequently harnessed for expressly augmentation for the multifunctional electronic devices, which provide us a new research domain known as ''MOFtronics'. The timeline of the contribution of key conducting MOF is illustrated in the **figure. 5**.

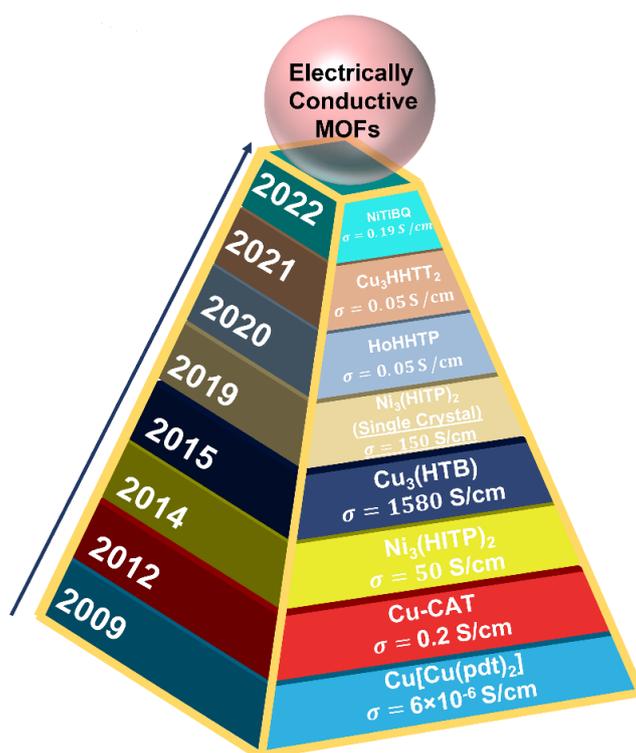

**Figure 5.** The timeline of representative conductive MOFs.

The conductivity of MOFs can be significantly enhanced by various means, including coupling metal sites to linkers, doping with redox-active guest molecules, and so on. Thanks to metal ions that have high-energy electrons or holes and organic linkers containing unpaired electrons or redox-active molecules to facilitate charge transfer between metal nodes. Furthermore, efficient spatial and energetic overlap between orbitals of suitable symmetry enhances charge transport, leading to increased manifold conductivity. So far, two physical models have been used to describe the charge transport in conductive MOFs: hopping transport and band theory (**figure 6**)[45].The former is a thermally triggered phenomena in which charge carriers (electrons, holes, or both) which are localised at specific sites, jump from one localised site to another sites, and conductivity increases as temperature rises (**figure 6b**)[46]. This mechanism of charge transport is typically found in MOFs with redox active centres, where charge transport occurs through ion diffusion through self-exchange reactions to balance the charge. An exponential equation describes the hopping mechanism's relationship to temperature, i.e., is given by equation (i).

$$\sigma = \sigma_0 \exp\left[-\left(\frac{T_0}{T}\right)^{1/d}\right] \qquad (i)$$

where $\sigma$, $T$, are conductivity, temperature respectively, whereas $\sigma_0$ and $T_0$ are constants specific to the material, and $d$ is the dimension of the sample.

In contrast, the latter exhibits continuous flow of electron because of strong interaction between the sites (**figure 6a**). It can also be thermally activated, and deactivated, conductivity will improve as temperature increases if the material possesses semiconductor properties, whereas conductivity will decrease as temperature rises for metallic samples.

The charge transport mechanisms (hopping and band theory) chemically can be accomplished with the 'through-bond' and 'through-space' approach[47]. The charge is transported via 'through-bond' , when strong covalent bonding of metal centres and organic linkers involves high spatial and energetic orbital overlap, resulting in the low band gap and high charge mobility. Such charge transport is more pronounced when soft and more electroactive linker atoms, such as nitrogen and sulphur-based linkers, coordinate with the metal d-orbital to a higher extent because they have well-matched energy levels and good orbital overlap, e.g., Fe-based MOF, consist of mixed-valent $Fe^{II/III}$ metal centres and azolate-based linker[48]. Whereas the charge is transported 'through space' because of non-covalent interactions between electroactive fragments, such as non-covalent $\pi$-$\pi$ interaction. MOFs composed of organic linker like triphenylene, trinaphthylene, hexaazatrinaphthylene, phthalocyanine, naphthalocyanine with ortho-substituted functional group (–OH, –$NH_2$, –SH, and –SeH) form $\pi$−$\pi$ interactions with one another, facilitate charge transport via through-space pathways.

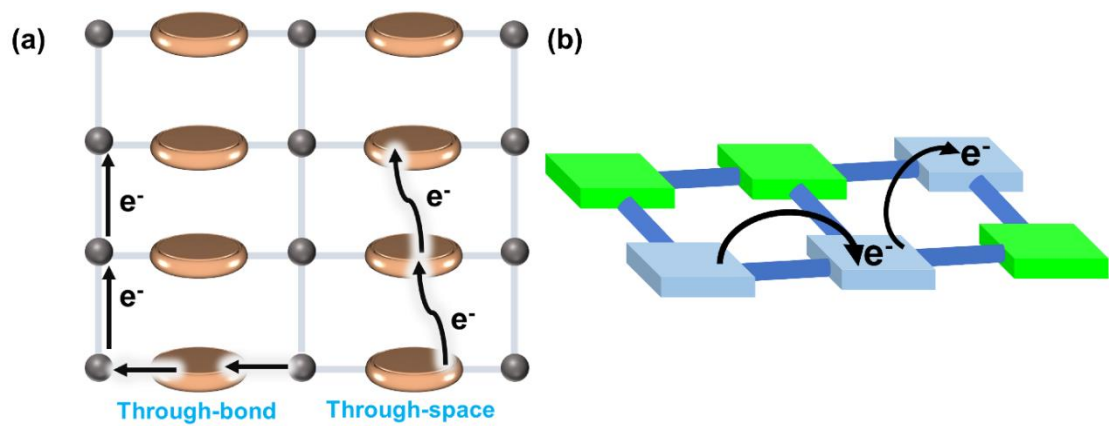

**Figure 6.** Schematic illustration of possible charge transport modes in conductive MOFs. (a) "through-bond", "through-space" and (b) charge hopping mechanism.

## 4. Electrical Characterization Techniques

The determination of a precise and accurate conductivity value is a necessity for the correct electrical assessment of MOFs in electronic devices such as batteries, supercapacitors, and field-effect transistors. The equation for electrical conductivity is as follows (ii)

$$\sigma = G\frac{l}{A} = \frac{I}{V} \times \frac{L}{A} \tag{ii}$$

Where $\sigma$, $G$, $l$, $A$ are electrical conductivity, conductance, the distance between the two probes/electrodes and cross-sectional area of the conductor, respectively. Resistance (R) of the sample is obtained using ohm's law. The equation for Ohm's law is as follows (iii)

$$V = IR \tag{iii}$$

V, I, and R are the voltage, current and resistance, respectively. Because the majority of conducting MOFs are semiconductors in the nature, their I-V plot deviates from Ohm's law, indicating that I varies non-linearly with V. As a result, apposite voltage or current is applied to the MOF sample to comply with Ohm's law.

MOF conductivity can be measured on single crystal and polycrystalline samples in the form of thin films or pellets. In polycrystalline samples, the grain boundary and sample anisotropy contribute to the conductivity in addition to the sample's inherent conductivity. Therefore, a single crystal is highly desirable to overcome these issues, because single crystals with no such contribution provide us the sample's inherent conductivity. However, growing a single crystal is a challenging and skilled operation. In addition, manipulating it for conductivity measurement is a tedious task because of its tiny size. As a result, in many cases, thin film or pellets are preferable over single crystal.

Conductivity is often measured using one of two methods: two probe or four probe[49]. The overall conductivity in the former is the sum of the conductivity of the substrate, contact, wire, and conductive paste, and the sample's contribution. As a result, this method works best with samples that have a low conductivity (high resistivity) value, i.e., whose resistance is insignificant in comparison to the wire and contact. The conductivity of MOFs is typically in the $10^{-3}$ S/cm range and resistance of wires and contacts is about 100 $\Omega$. Therefore, this method is suitable for electrical characterization of such high-resistive materials. To assess the conductivity of the sample impeccable contacts are made by employing conductive metal such as Ag/Au/Cu/Al and conductive adhesive like carbon paste, gold, or silver paint. For instance, Dinca and co-workers measured the conductivity of the $Cd_2(TTFTB)$ MOF by fabricating a device, in which they first prepared pellet followed by using conducting carbon paste and gold wires, to make two contacts.[50] This method involves applying a known voltage or current to the sample and calculating the resistance value using Ohm's law (**figure 7**).

Recently, high conducting MOFs have been reported that have conductivity comparable to that of wire and contact. Therefore, four-probe methods for measuring its conductance become quite relevant, as the four-probe method has a configuration that exclusively measures the intrinsic conductivity of the sample excluding the conductivity of the contact and the wire. In this method the four probes are placed in a straight line at equal distances, which simplifies the calculations and makes the probes easy to set up. Current is applied through the outer two probes, while the potential is measured through the inner two probes. The voltameter has a resistance ranging from $10^{12}$ to $10^{17}$ $\Omega$, and because it is connected parallel to the sample, most of the current passes through the sample due to the voltameter's high resistance. As a result, even the sample has a resistance in the range of $10^9$, which can be estimated quite accurately using the four-probe method. The precise determination of conductivity is dependent on the probe making flawless contact with the sample; in comparison to the four-probe method, the two-probe method is considerably easier to connect with the sample. To confine the electric conductivity of MOFs, two probe methods are usually chosen over four-probe methods.

The four-probe method works reasonably well when the film thickness is infinite, while the Van der Pauw method is used when the sample is of relatively small thickness or irregular shape (Figure 7d)[51]. Because it is less sensitive to the shape of the sample, the Van der Pauw is one of four probe methods used to measure conductivity of irregular shape samples. Four probes are arranged squarely around the perimeter of a sample, where current is applied through two probes on one edge, while two probes on the opposing edge measure voltage. The resistivity along the sensing direction is estimated by a linear array in the four-probe method, whereas in the van der pauw method, two resistances are calculated, one along the applied current and the other along the measured voltage, and then the average of the two is determined.

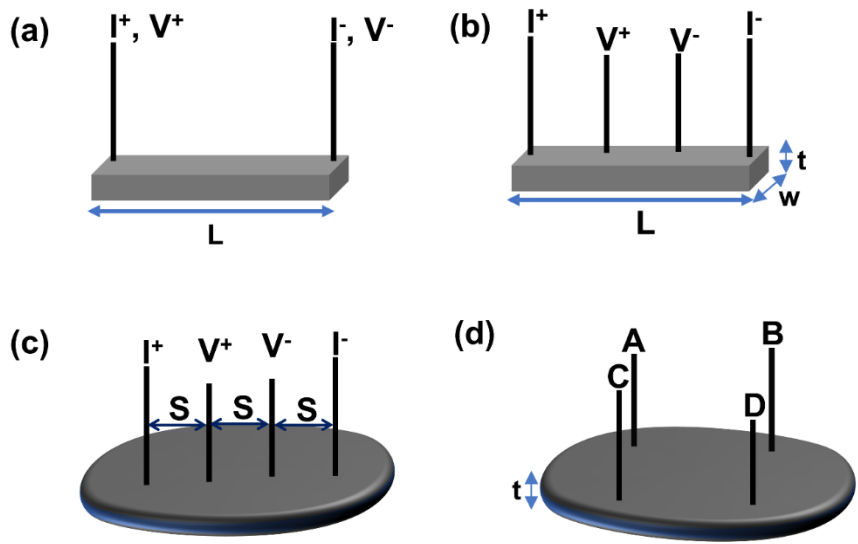

**Figure 7.** Schematic diagram of method used to measure electrical conductivity of MOF (a) Two-contact method (b) Four contact method (c) Four probe method (d) Van der paw method. Where I, V, L, W, and T correspond to current, voltage, length, width, and thickness, respectively.

## 5. External stimuli-driven charge transport in the MOF

The applied physical stimuli (for instance, electric or magnetic field, light, temperature, and so on) have been played a significant role in the modulation of electronic properties of MOFs. For example, McGrail and co-workers reported the mechanism for which a MOF composed of TCNQ (7,7,8,8-tetracyanoqunidodimethane) linker and $Cu^{2+}$ metal node can reversibly change the shape under an applied potential into either a conducting phase (conductivity value $4.8 \times 10^{-3}$ S $cm^{-1}$) or an insulating phase (conductivity value $5.8 \times 10^{-7}$ S $cm^{-1}$) [52]. Herein, for charge transport investigation, a device was fabricated, by placing a thin film of Cu(TCNQ) in between the Al foil and Cu contact (**Figure 8**). The corresponding I-V curve illustrates an almost negligible current from 0 to 1 V, designated as an ''off'' insulating state. After that, the current of Cu(TCNQ) film moves to the ''on'' conducting state at approximately 4.5 V. This phenomenon clearly indicates that upon applying the bias, the potential of insulating phase (impedance about 100 kOhm) was converted into the conducting phase (impedance 1500 ohm). In the conducting phase, a layer structure of Cu(TCNQ) was formed in such a fashion that bettered the interaction between the d orbitals of metal with the p orbitals of the cyano-bridge, resulting in the considerable enhancement in the electrical conductivity [53].

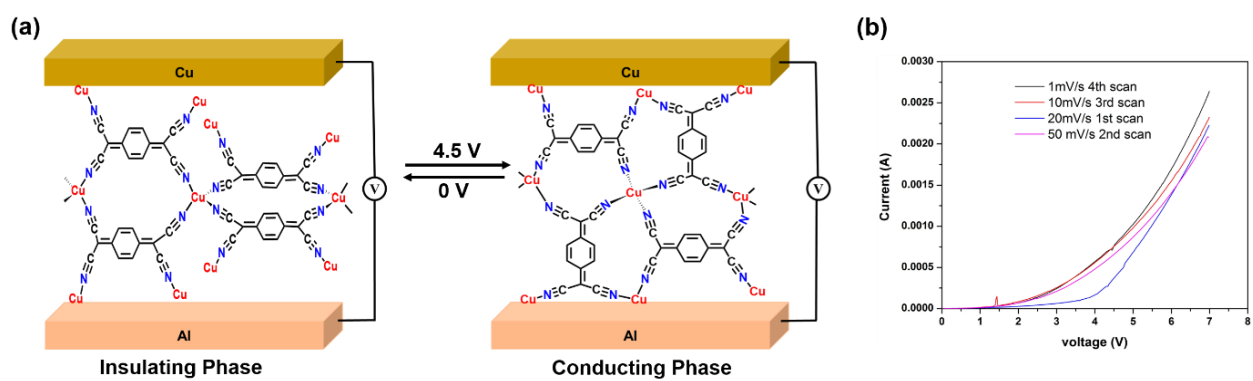

**Figure 8.** (a). Schematic illustration of reversibly switching of a device (Al/Cu(TCNQ)/Cu) between conducting and insulating phase upon applying potential 4.5 V and 0 V respectively. (b) I-V curve recorded between 0 to +6 V at different scan rates presenting bistable switching behavior of a 10 mm-thick Cu(TCNQ) phase II film. Reproduced with permission from ref. [52], copyright 2015, Science.

Photo-driven charge transport studies are another exciting field in molecular electronics owing to their intriguing application like photoswitches, photosensors, photoelectrode, photocatalysis, etc. However, the absence of photosensitive linkers in MOF moieties precludes it from integrating photo-driven electronic devices. For the first time, Shustova and co-workers combined photosensitive linkers spiropyran and diarylethene on MOF for electronic structure modulation [54]. Herein, upon UV (365 nm) irradiation on spiropyran linker, a charge-separated merocyanine was formed, intensifying charge-hopping rates (**Figure 9**). Thereby, a significant jump in electrical conductivity (1.2 times) was observed. Apart from that, upon UV irradiation on the diarylethene linker closed conjugated pathway was formed. As a result, a prominent electrical conductivity (($9.5 \pm 2.1) \times 10^{-7}$ Scm$^{-1}$) value was obtained. An open non-conjugated and low conducting ($2.9 \pm 0.67) \times 10^{-6}$ Scm$^{-1}$ pathway was restored, on further irradiation with visible light,. Furthermore, the enhancement in electrical conductivity was successfully established by using the breadboard circuit setup. Thus, the photosensitive linker facilitates the charge transport by the impact of light-stimuli in the MOF scaffold. This accomplishment opened the window of enormous possibilities to explore MOF into the realm of photo-driven molecular electronic devices.

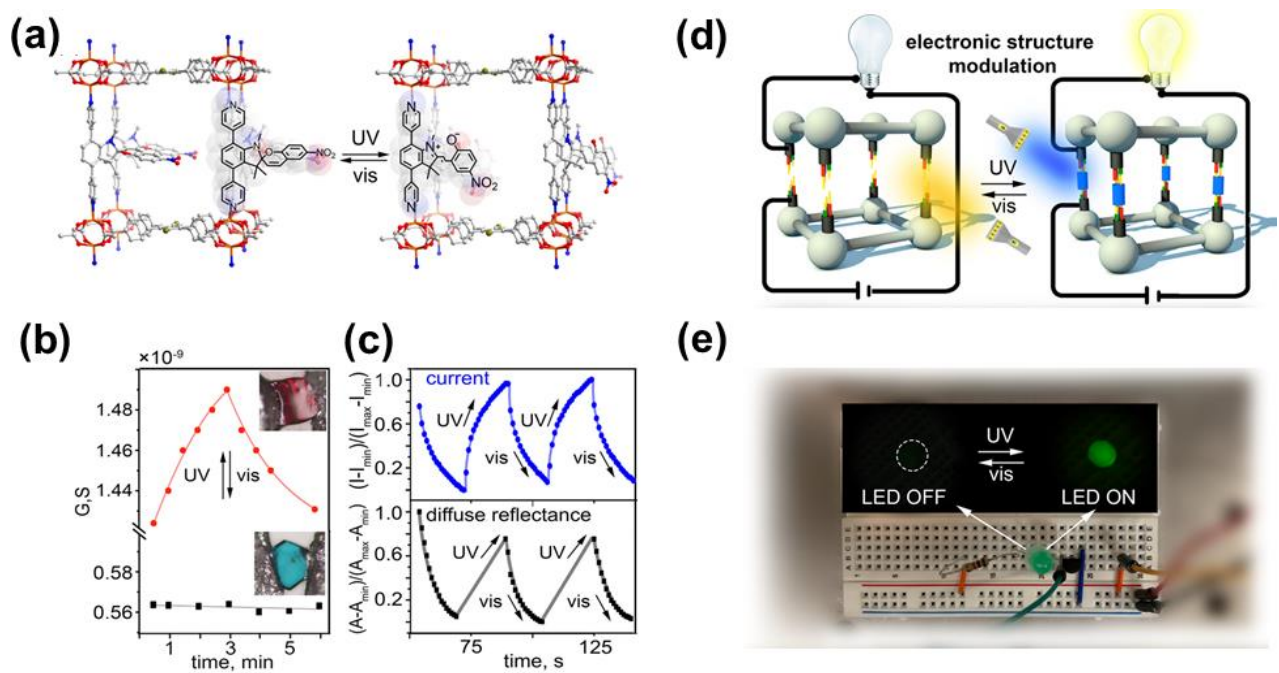

**Figure 9.** (a) The formation of charge-separated merocyanine upon UV (365 nm) irradiation, neutral spiropyran was restored upon visible-light (590 nm) irradiation. (b) Variation of electrical conductance with UV-visible irradiation (inset illustrate single crystal set up) (c) Normalized reversible photoisomerization cycle of current (top) and absorption (bottom) with a variation of UV-visible irradiation. (d) Schematic illustration of the MOF as a circuit element. (e) Demonstration of enhancement in MOF Electrical conductivities by using breadboard setup. Reproduced with permission from ref. [54], copyright 2019, American Chemical Society.

A solvent can also modulate the electronic properties of MOFs, either changing the chemical composition, geometry, or through modifying the host-guest interactions [55,56]. For instance, the coordination of the *N, N*-dimethylformamide (DMF) molecules to the unsaturated Fe center of Fe$_2$(DSBDC) (DSBDC = 2,5-disulfidobenzene-1,4-dicarboxylate) bolster the electrical conductivity by a thousand-fold at room temperature [57]. The electrical conductivity was measured using the paste-wire method, in which the pellet was sandwiched between the carbon paste and gold wire (**Figure 7**). The DMF-soaked MOF showed improved electrical conductivity ($10^{-6}$ Scm$^{-1}$). In contrast, on the removal of unbound DMF, electrical conductivity was observed to be less ($10^{-7}$ Scm$^{-1}$), and on complete removal of DMF, the electrical conductivity was even less than the preceding one ($10^{-9}$ S cm$^{-1}$). The increase in the electrical conductivity was attributed to the fractional electron transfer from Fe to DMF. Theoretical calculations revealed the ionization potential of DMF is lower than the work function of the Fe$_2$(DSBDC) by 1 eV and virtually electron transfer takes place when DMF binds to the activated Fe$_2$(DSBDC) levels of DMF.

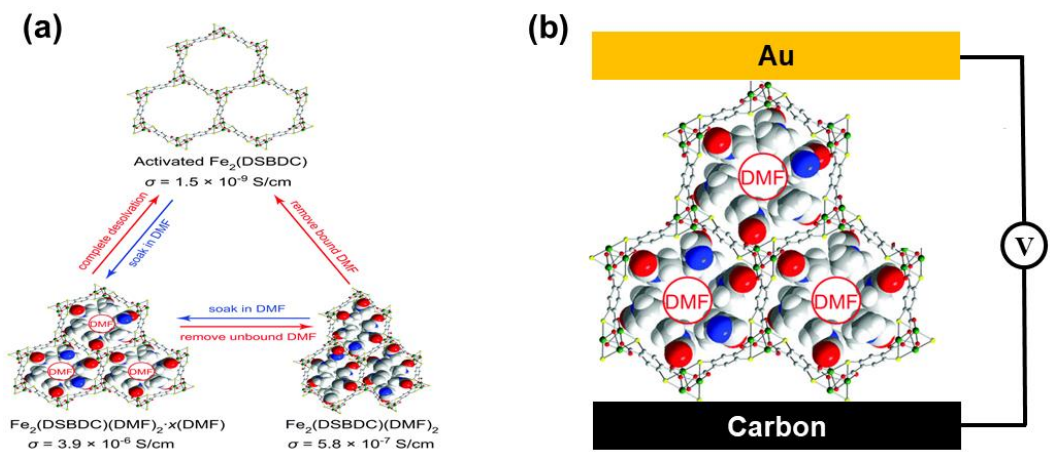

**Figure 10.** (a) The variation of electrical conductivity and structural change of (Fe$_2$(DSBDC) MOF upon solvation with DMF solvent. (b) Schematic diagram for Au/Fe$_2$(DSBDC) (DMF)$_2$.X(DMF) /Carbon molecular junctions. Reproduced with permission from ref. [57], copyright 2018, The Royal Society of Chemistry.

It is important to be mentioned that Robin and Day earlier classified mixed-valence materials into three different categories [58]. Category I is based on the geometry and coordination numbers of the central metal ions where, the electron density of metal-ligands has mostly been localized and considered insulators. Category III materials are characterized by identical stereochemistry at metal centers with completely delocalized, i.e., indistinguishable, oxidation states. These materials behave as two- or three-dimensional conductors. Class II occupies an intermediate position between class I and III; i.e., these compounds are characterized by similar stereo chemistries (differing from one another by slight distortions) at metal centres with (more or less) localized, distinguishable, oxidation states. For these materials, which are semiconductors, the establishment of a superlattice is foreseen. Next sections, we will discuss the role of diverse dopants in the enhancement of electrical conductivity.

# 6. Charge transport in the doped-MOFs

Usually, classic mainstream MOFs are poor conductors of electricity [59]. One of the possible reasons is the dimensions of porosity and less abundant exposed active sites. [60,61]. However, pore size is large enough to accommodate small electroactive molecules, such as ferrocene, TCNQ, conducting polymer, anthracene, Iodine, etc. Consequently, the MOF community has successfully turned this challenge into an opportunity by post-synthesis doping, resulting in the excellent augmentation in terms of charge transmission for the applied research [15,62–64]. To date, several MOFs have been transformed into a conductor by exploiting its porosity. For instance, Farha and Co-worker doped NU-1000, which consists of the pyrene-based linker and Hexa-Zirconium nodes with Nickel (IV) bis(dicarbollide) (NiCB)[65].

The electrical conductivity was measured by drop-casting the NU-1000, NiCB, and NU-1000@NiCB on the interdigitated electrodes (IDEs) consist of platinum fingers. NU-1000@NiCB exhibited electrical conductivity upto $2.7 \times 10^{-7}$ S/cm **(Figure 11a)**. The increase in the electrical conductivity was evidenced by I-V measurement. In the correspoding I-V curve, zero slopes were obtained for both NU-1000, and NiCB, whereas a non-zero slope was obtained for NU-1000@NiCB **(Figure 11b).** The remarkable enhancement of electrical conductivity was attributed to the donor-acceptor charge transfer between the pyrene-based linker and NiCB guest.

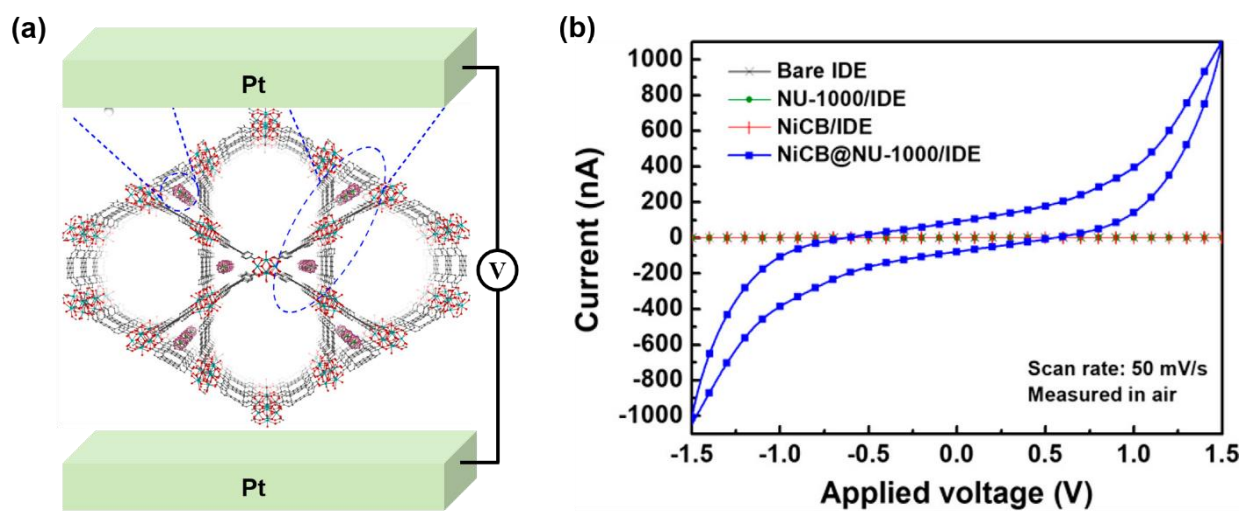

**Figure 11:** (a) Schematic of molecular junction Pt/NU-1000/Pt. (b) I-V curves show zero slope for bare IDE, NU-1000/IDE, NiCB/IDE, and non-zero slope for NiCB@NU-1000/IDE. Reproduced with permission from ref. [65], copyright 2018, American Chemical Society.

Conducting polymer fibers embedded on MOFs have recently showcased a drastic increase in the electrical conductivity [66,67]. For instance, UIO-66 was an insulator owing to the lower electrical conductivity ($\sim 10^{-8}$ S cm$^{-1}$). However, when UIO-66 was loaded with Poly-Pyrrole (PPy) and Poly 3,4-ethylenedioxythiophine (PEDOT), the electrical conductivity of UIO-66@PDY and UIO-66@PEDOT was obtained to be $\sim 2 \times 10^{-2}$ S cm$^{-1}$ and $10^{-3}$ S cm$^{-1}$ for four-probe and $\sim 7 \times 10^{-2}$ S cm$^{-1}$ and $\sim 4.5 \times 10^{-4}$ S cm$^{-1}$ for two probe DC techniques, respectively **(Figure 9)**. The results advocate the formation of a single chain of conjugative polymer inside framework voids providing the pathway for the effective charge transport across the hybrid nanomaterial [68].

The tuning of the bandgap is another important parameter to boost the electrical conductivity of the MOF. [69–72]. Recently, Ogihara et al. tuned the bandgap of 2,6-naphthalene dicarboxylate dilithium via Li-intercalation. [73]. Due to Li-intercalation, the bandgap was reduced to 0.9 eV and presumably followed the hopping pathway for electronic charge transmission. Besides, the electrical conductivity of intercalated MOFs (IMOFs) was found to be reversibly dependent on temperature. It exhibited the conductivity up to 200 °C and then became electrically inactive. **(Figure 12 a-b)**. Thus, it can be employed as a prototype switch for protection at high temperatures in electronic devices. Furthermore, the charge transport properties were investigated by electrochemical impedance spectroscopy (EIS) studies. The EIS measurements were performed by applying 500 mV amplitude in the frequency ranges from 100 kHz to 1 Hz at OCP. Bode plot showed resistance for Li-intercalated MOF four order less than pristine MOF at the low-frequency regime (**Figure 12c**). In the corresponding Nyquist plot, a small semi-circle appeared for Li-intercalated MOF, representing the contribution of ions and electrons towards lower resistance (**Figure 12d**). In contrast, for pristine MOF, a straight line was found that render mainly contribution of ionic conductance. Thus, EIS studies provide insight into the charge transport phenomenon, by revealing the individual contribution of charge carriers (ions and electrons)[74].

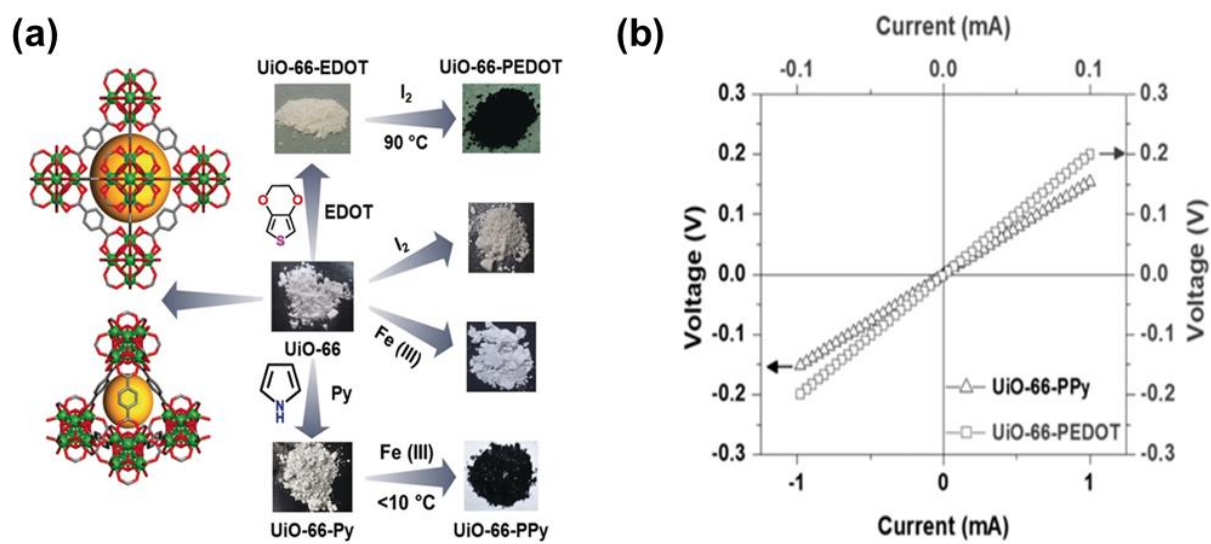

**Figure 12.** (a) Schematic illustration of the synthesis of UiO-66-PPY and UiO-66-PEDOT, where UiO-66-PPY was synthesized by first loaded with Py followed oxidation with $FeCl_3$ and UiO-66-PEDOT was synthesized by first loaded with EDOT followed by oxidation with $I_2$. (b) Comparison of I-V curve of UiO-66-PPY and UiO-66-PEDOT. Reproduced with permission from ref. [68], copyright 2020, Wiley-VCH.

In this review, we have already discussed the synthesis of HKUST-1, but insight into the charge-transport mechanism remains to be addressed. In this context, Liu *et al.* synthesized an array of thin films of HKUST-1 MOF by varying the number of cycles in the LPE method, yielding thin films of thickness, d = 45, 58, and 71 nm (approximately) corresponding to the different number of cycles [24]. Subsequently, thin films were loaded with ferrocene (fc) molecules. The tunneling molecular junction with a stacking configuration Au/ CMMT SAM /HKUST-1/HDT SAM/ Hg was made to unravel the charge transport phenomenon by varying thicknesses (**Figure 13a**). (**Figure 13b-c**) shows the corresponding I-V curve featuring highest current density of doped HKUST-1 SURMOF and the linear decreases of current density with the film thickness producing a low decay constant ($\beta \approx 0.006$ Å$^{-1}$). Moreover, the linear increases of resistance with thickness were also speculated (**Figure 13c inset**) which revealed that charge transport is taking place via an incoherent charge-hopping mechanism. Additionally, the electrical conductivity of Fc doped HKUST-1 was found to be higher than that of the pristine one, ascribing the reduction of the charge injection barrier due to electronic coupling between the metal nodes.

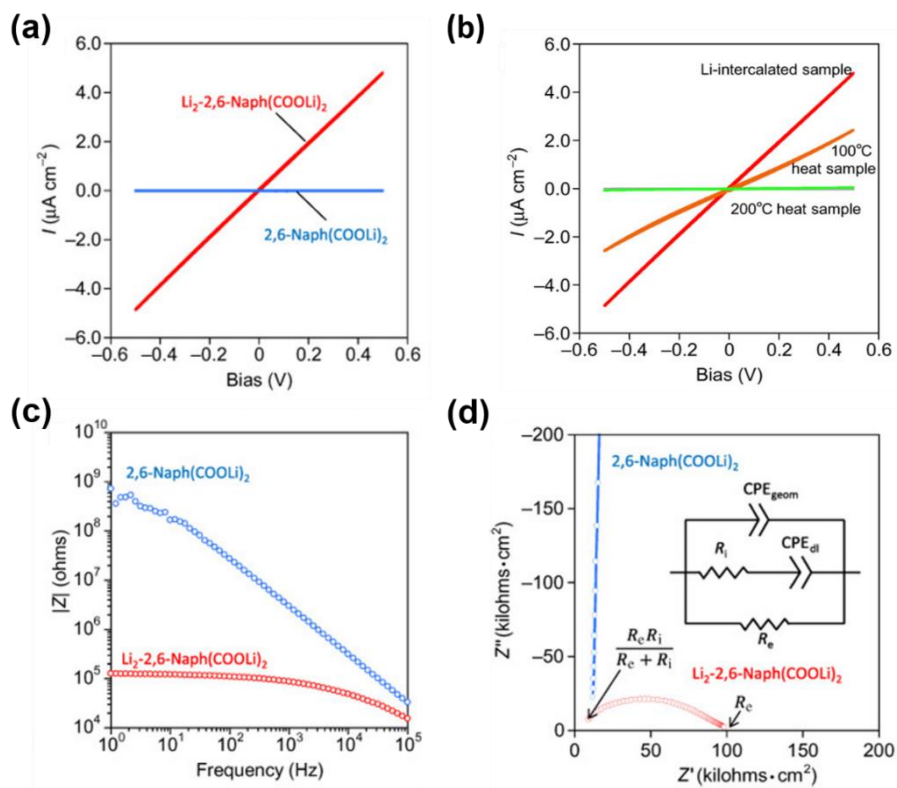

**Figure 13**. (a) I-V plot shows pristine MOF as an insulator and Li-intercalated MOF as a conductor. (b). Shows decline of conductivity with temperature, at 200°C it turned to insulator. (c) Bode and (d) Nyquist plot, respectively. The inset in Figure (d) represents the corresponding equivalent circuit. Reproduced with permission from ref. [73], copyright 2017, Science.

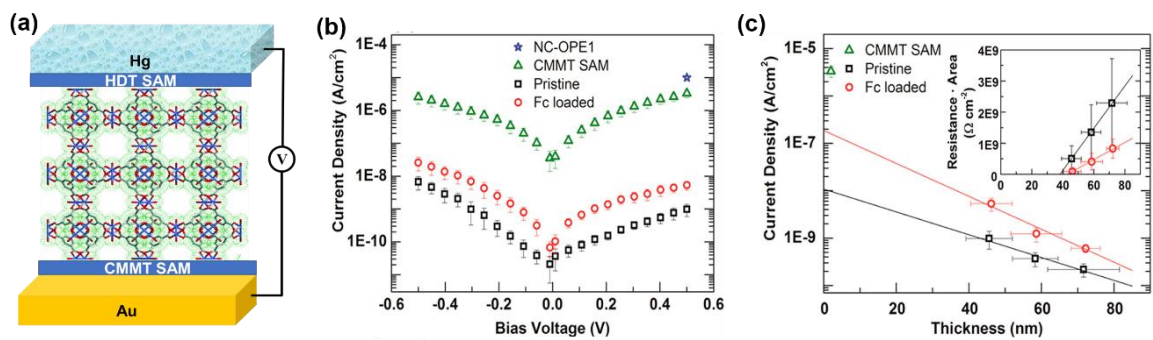

**Figure 14.** (a) Schematic illustration of device Au/ CMMT SAM / /HKUST-1/HDT SAM/ Hg. The device was used for thickness-dependent charge-transport studies. (b) Semi log J-V curve illustrates current density of doped HKUST-1 SURMOF is higher than that of pristine and high electrical conductivity of CMMT SAM shows defect-free molecular junction. (c) Linear decreases of J with thickness and linear increase of resistance with a thickness (inset), respectively. Reproduced with permission from ref. [24], copyright 2017, American Chemical Society.

For a profound understanding of the charge transport in MOFs, a defect-free large-area thin film of MOFs are another prerequisite item. To date, numerous methods such as solvothermal, LPE (vide supra), and gel-layer growth have been explored for the thin-film synthesis of MOFs. Nonetheless, these methods have not proved to be so efficient for producing pinhole-free, crystalline, high throughput, and large-area thin films. In this context, recently Jung et al. synthesized a thin film of HKUST-1 MOF with a large area, and high crystallinity by employing the solution shearing method [75]. The thin film was photosynthetically doped with TCNQ molecule, resulting in a significant increment of the electrical conductivity up to seven orders of magnitude. The electrical conductivity was measured with the variation of the soaking time of the TCNQ molecule. An increase in the electrical conductivity was found with an increase in the soaking time, as the TCNQ molecule gets more exposure to bond with the open metal centres. Furthermore, the conductivity of the thin film was found to be higher than that of the pellets. The authors concluded that since the synthesized thin film is free from defects such as pinholes, and grain boundaries, its conductivity is much higher than that of pellets.

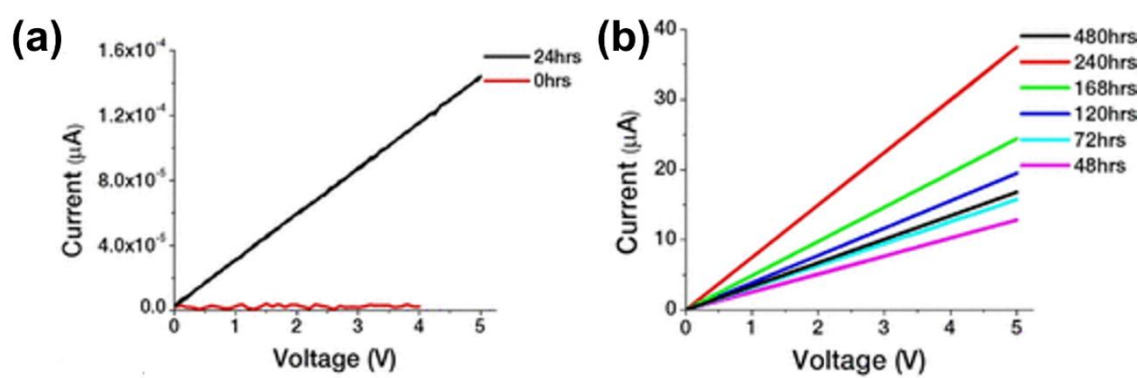

**Figure 15.** (a,b) The current V/S voltage plot of TCNQ@HKUST-1 shows the linear variation of current with respect to voltage over the time period of TCNQ soaking. Reproduced with permission from ref.[75], copyright 2021, American Chemical Society.

Another approach to invoke the host-guest chemistry for electrical conduction is introducing *in situ* polymerization inside the restricted space of redox-active polymer. Towards that direction, Ballav and co-workers recently combined redox-active Pyrrole (Py) into the nanochannels of $[Cd(NDC)_{0.5}(PCA)]G$, further on treatment with Iodine ($I_2$), the polymer of pyrrole was formed, resulting in the incredible enhancement in the conductivity [76]. **Figure 12a** shows Field-emission electron microscopy (FESEM) images advocate the same morphological patterns of $[Cd(NDC)0.5(PCA)]$ and $[Cd(NDC)0.5(PCA)]PPy$. Energy dispersive X-ray spectroscopy (EDXS) indicates the presence of nonstoichiometric and insignificant amount of $I_2$ corroborating its role as an initiator for Py oxidative polymerization [77]. $[Cd(NDC)0.5(PCA)]$ exhibited conductivity upto ~$10^{-12}$ S cm$^{-1}$. However, on incorporation with PPy, electrical conductivity was raised to ~$10^{-3}$ S cm$^{-1}$ (**Figure 16b**). This billion-fold increment of electrical conductivity was ascribed to the formation of highly oriented conducting PPy inside 1-D nanochannel of 3D MOF. More specifically, the non-covalent π-π/ N-H-π type interaction between polypyrrole and 3-D MOF providing percolating paths has been considered the leading cause behind such a notable enhancement. Also, they revealed the formation of Polaron/bipolar chain by getting intense UV-vis spectrum at 620 nm corresponding to the small value of $E_g$ electronic state and thus indicating the unusual enhancement for conductivity within the relatively doped state [76]. In this context, Dincă and co-workers have summarized some unifying results of enhanced conductivity achieved by incorporating polymer chains into MOFs by assembling monomers into the channels and mentioned the activation chemistry using I$_2$ to influence polymerization [78].

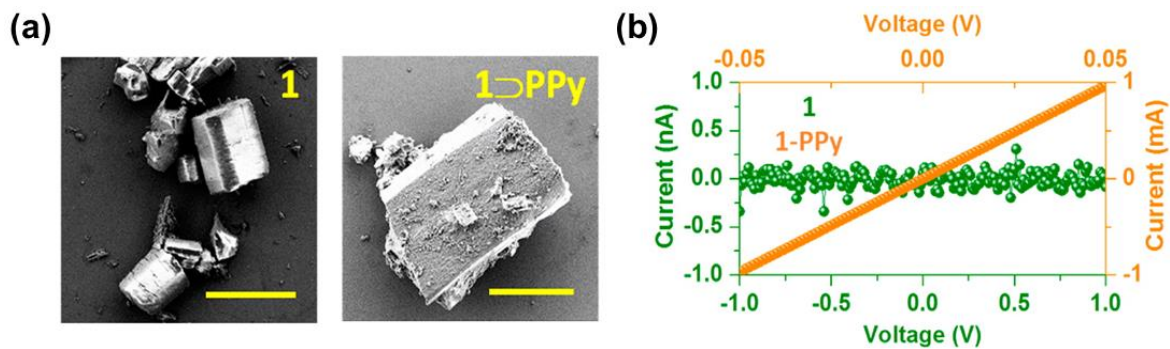

**Figure 16.** (a) FESEM images of [Cd(NDC)$_{0.5}$(PCA)] and [Cd(NDC)$_{0.5}$(PCA)]PPy (b) I-V Plot of [Cd(NDC)$_{0.5}$(PCA)](green) and [Cd(NDC)$_{0.5}$(PCA)]PPy(Orange). Reproduced with permission from ref. [76], copyright 2016, American Chemical Society.

Although several research groups have extensively studied the effect of halogen doping on the electrical conductivity of MOFs. However, the effect of stoichiometry ratios, structure-property relationships and conductive carriers of MOFs has yet to be addressed. In this context, recently Gupta et al. for the first time investigated the optical, thermoelectric, and semiconducting properties of the MOF, Cu[Cu(pdt)$_2$] (pdt =2,3-pyrazinedithiolate) by infiltration with vapour of bromine with varying stoichiometry ratios[79]. Due to doping, Cu[Cu(pdt)$_2$] is partially oxidised, forming a mixed valency Br$_x$@ Cu$^{II}$[Cu$^{II}$(pdt)$_2$]$_{1-x}$[Cu$^{III}$(pdt)$_2$]$_x$. The mixed valanced compound facilitates the charge transport via hopping mechanism, resulting in a tenfold increase in electrical conductivity. Furthermore, for x (Br fraction)> 0.5, the pristine MOF switches from p-type to n-type semiconductor. The pristine MOF exist in Cu$^{II}$[Cu$^{II}$(pdt)$_2$] state, which can be hardly reduced but can be readily oxidized to [Cu$^{III}$(pdt)$_2$]$^-$. It is therefore recommended the hole charge carrier is thermally generated and exhibited as a p-type semiconductor. In contrast, the doped MOF exists in the [Cu$^{III}$(pdt)$_2$]$^-$state, which can be easily reduced to [Cu$^{II}$(pdt)$_2$]$^{2-}$ but not oxidised, suggesting that the thermally generated charge carriers are electron. Therefore, it can act as an n-type semiconductor.

Wenzel and co-workers revealed the reason behind the high conductivity value of MOF loaded on TCNQ. Although it remained challenging to resolve the predictable transport mechanisms since the obtained experimental data neither showed simple hopping nor band transport models [80]. However, the theoretical studies agree with the experimental data to explain that the observed conductivity could be due to an extended hopping transport model, including virtual jumps within localized MOF states or molecular super exchange [17]. The temperature dependence study demonstrated that band transport and hopping mechanisms are fundamentally different. In the band transport mechanism, the increasing temperature decreases the conductivity owing to augmented electron (or hole)-phonon scattering. On the other hand, increasing conductivity with the temperature usually assists towards the hopping type of conduction path. Although at nearby room temperature, band conduction is depicted for an intrinsic semiconducting MOF with a bandgap of almost 0.3 eV or even lower [81]. Pathak *et al.* recently reported high electrical conductivity (10.96 S cm$^{-1}$), and small bandgap (1.34 eV) of single crystals Cu-based MOF {[Cu$_2$(6- Hmna)(6-mn)]·NH$_4$}$_n$ (where, 1, 6-Hmna = 6-mercaptonicotinic acid, 6-mn = 6-mercaptonicotinate), consist with a two-dimensional copper–sulfur plane [82]. They explained that expanding the (-Cu-S-)$_n$ chain to (-Cu-S-)$_n$ planes permits the charge transfer to occur along with two directions, resulting in high electrical conductivity. Consequently, this approach can be denoted as a model for designing conductive MOFs in crystalline form for applications in other energy storage fields like a supercapacitor. In addition, thickness and temperature-dependent charge transport studies were performed. With an increase in the thickness, a considerable decrease in the electrical conductivity was observed due to crystal defect and the semiconducting property nature of MOF. However, the conductivity was increased with an increase in the temperature (**Figure 17b).**

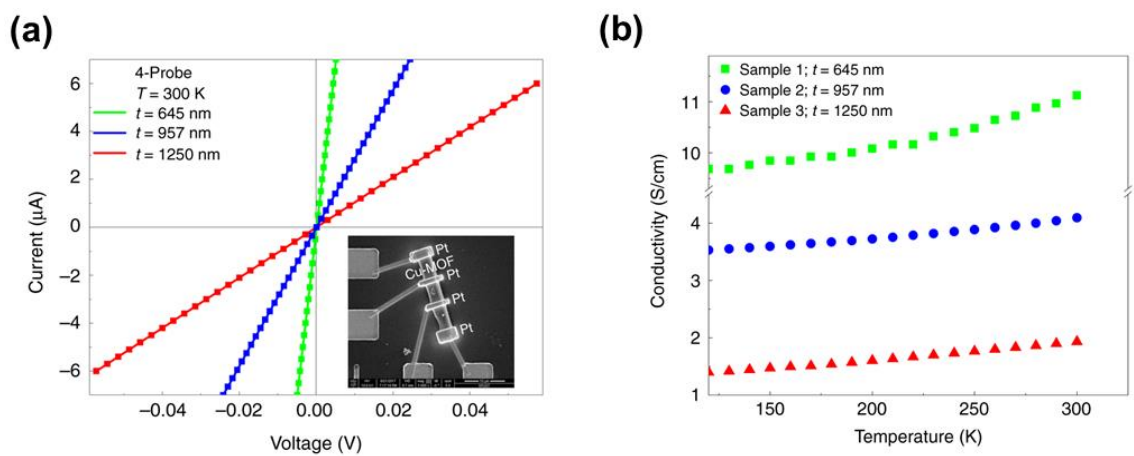

**Figure 17:** (a) Comparison of I-V curves of a single crystal of different thickness, measured using four-probe methods. (b) Illustrate enhancement in the electrical conductivity with temperature. Reproduced with permission from ref. [82], copyright 2019, Springer Nature.

Microporous materials introduce a large family of solids having distant applications appropriate to the reachable exposed sites and chemical functionality. For example, Alkordi and co-workers revealed that enhanced electrical conductivity and maintained microporosity could simply and controllably be reached through well-organized compositing of monolithic forms of the crystalline hybrid MOF with conductive components like graphene [83]. The electrical conductivity was found to be high (in the range of $7.6 \times 10^{-6}$ S m$^{-1}$ to $6.4 \times 10^{-1}$ S m$^{-1}$) and persisted comparative with that of graphene content. Moreover, the approach offered a solution to a mouldable MOF@G composite that circumvents significant challenges encountered in composites containing graphene or CNTs, namely processing contests that prohibited full utilization and exploitation of such composites [84]. The electrical conductivity of MOFs can also be substantially enhanced by the modulation of their chemical structure. For instance, most recently Zhu and co-workers fabricated three electrically conductive MOFs by employing two strategies (i) by tuning the oxidation state of the ligand and (ii) using dopant [85]. The authors synthesized MOF MnHHB (HHB = hexahydroxybenzene) by altering the oxidation state of MOF MnTHBQ (THBQ =tetrahydroxy-1,4-benzoquinone) and upon doping with metal ion (Rb⁺ or Cs⁺), resulting in two new MOFs, MnRbTHBQ and MnCsTHBQ, respectively. Remarkably, the electrical conductivity of MOF MnHHB was found to be higher than that of MnRbTHBQ and MnCsTHBQ. The π-π stacking distance for the MnRbTHBQ and MnCsTHBQ was ~3Å, whereas ~2.93Å for the MnHHB. The authors concluded a 5% lower π-π stacking interaction, inducing the closer distance of benzene moieties ligand to each other, which results in a hundred-fold enhancement in the electrical conductivity. Furthermore, temperature-dependent studies denoted the obtained band gap positioned in the semiconductor range.

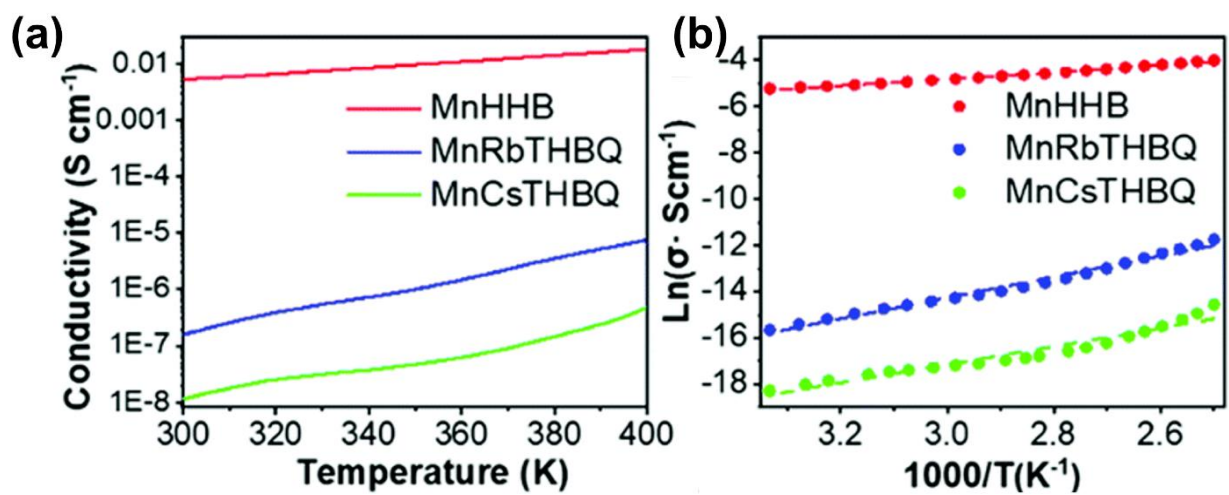

**Figure 18:** (a) The conductivity v/s temperature plots show the highest conductivity for MnHHB, then MnRbTHBQ and the lowest for MnCsTHBQ. (B) From the Plots of ln s(T) v/s 1000/T band gap (Ea) was obtained 0.128, 0.39 and 0.34 eV, MnHHB, MnRbTHBQ and MnCsTHBQ respectively. Reproduced with permission from ref.[85], copyright 2022, The Royal Society of Chemistry.

## 7. Charge transport studies in 2D and 3D Metal-Organic Framework

No doubt, 2D atomic crystals from the last several years have shown wide application for the budding field like energy, environment, and medical management [86–88]. Compared to other inorganic 2D materials, the formation of 2D MOFs through self-assembly of perceptible organic linkers and preferred metallic nodes produces numerous structural motifs, particularly intrinsically conductive material similar to traditional MOF [89–92]. After the first electrically conducting MOF reported in 2010 [93], last decades have witnessed a gigantic evolution of conducting MOFs. The coordination of metal centres with highly conjugated amino and thiol phenylene/benzene-based linkers such as 2,3,6,7,10,11-hexahydroxytriphenylene (HHTP),2,3,6,7,10,11-hexaiminotriphenylene (HITP), 1,2,3,4,5,6-hexahydroxybenzene (HHB) mention to few, yields 2D MOF that is similar in extended 2D graphene-like honeycomb grids [94–96]. In a recent review, Li *et al.* summarized the current progress of stimuli response 2D MOFs [97]. They have emphasized the future outlooks of various fast stimuli responses for the intrinsic structure, high porosity, exposed metal sites, high surface area, and large electrical conductivity associated with 2D MOFs.

## 7.1 Transition metal-based 2D MOF

The M-CAT-1(metal-catecholate) family exhibit strong charge delocalization over the plane due to the apparent energetic overlap between the metal nodes and the oxidized form of the organic linker. The alliance of in-plane conductivity and layered structure originates these systems an ideal platform to generate ultrathin films through the controlled transfer of pre-formed layers that reserve the electronic features of the bulk. However, the conductivity of thin film of M-CAT-1 MOFs has been observed lower (up to three orders) than the single crystal. Therefore, it is highly desired to synthesis crystalline and well-oriented thin films to overcome the low conductivity issue. Toward that goal, the Medina group recently, synthesized well-oriented thin films of the M-CAT-1 series, comprising of HHTP linker and metal nodes such as $Ni^{2+}$, $Co^{2+}$, and $Cu^{2+}$ on conductive surfaces using versatile vapor-assisted conversion (VAC) technique [98]. In this technique, a solvent containing an organic linker and metal ions is placed in a tightly closed vessel and the substrate onto a glass spacer. Afterward, a closed vessel is placed in a preheated oven for a certain amount of time, resulting in a highly oriented and compact, thin film having high surface coverage. **(Figure 19).** Here, the acid modulator immobilizing on a substrate plays a key role in defining thin-film orientation.

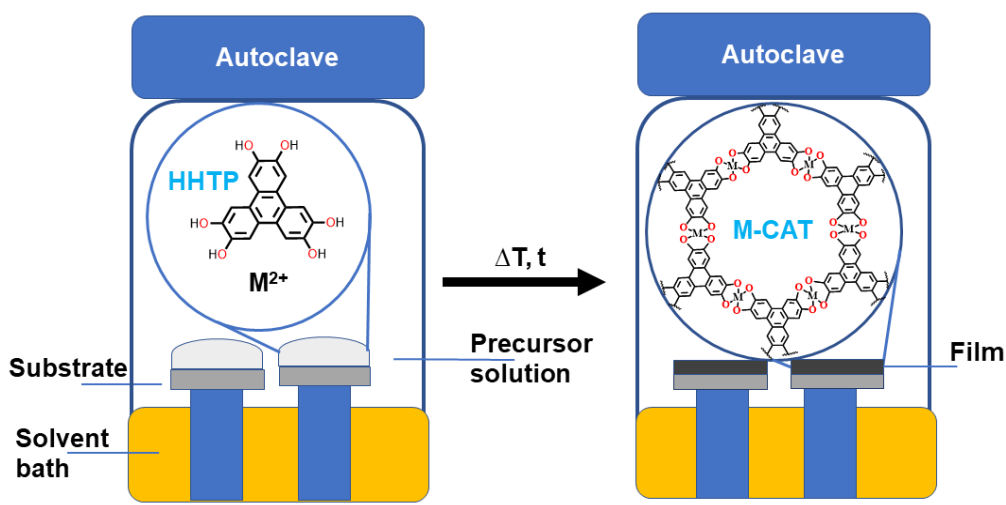

**Figure 19:** Schematic illustration of the VAC setup.

In addition, a solar cell device was fabricated by sandwiching a thin film of Ni-CAT-1 between ITO and Al substrate. **Figure 20a** pictorially represents the device configuration. Interestingly, thin films of Ni- and Co-CAT-1 grown on the quartz substrate exhibited electrical conductivity $1.1 \times 10^{-3}$ S cm$^{-1}$ and $3.3 \times 10^{-3}$ S cm$^{-1}$ respectively **Figure 20b**. The devices showed diode type behaviour under AM 1.5G light illumination which is reflected in the corresponding current–voltage curve in **Figure 20c**. The results depicted the layers of the molecular stack in the oriented thin film can serve as a better conduction path at a high surface coverage for the photogenerated charge carriers.

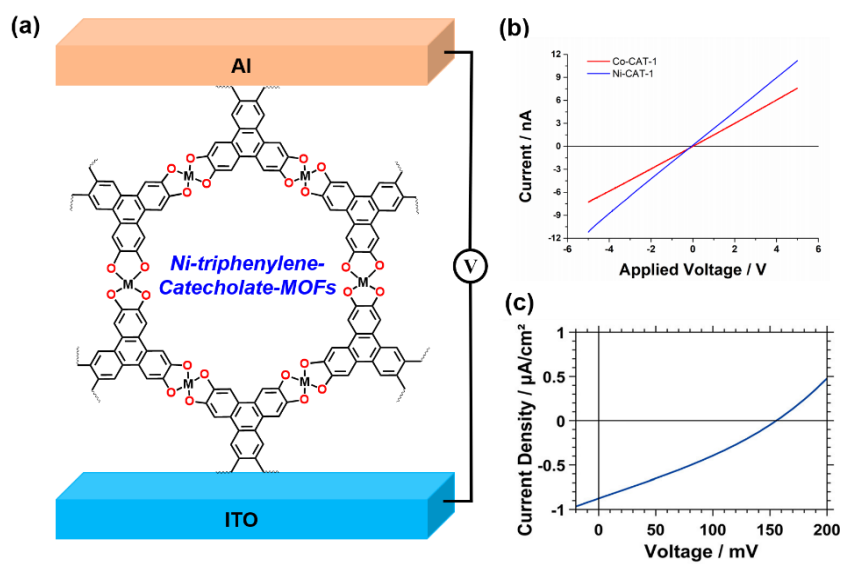

**Figure 20:** (a) schematic illustration of Solar cell device having configuration Al/ Ni-CAT-1/ITO. (b) I-V curve of Co-CAT-1 and Ni-CAT-1 thin film grown on a quartz substrate. (c) I-V plot of solar cell device representing diode characteristics. Reproduced with permission from ref. [98], copyright 2019, American Chemical Society.

Even though the research team have successfully synthesized the thin film of Co-CAT-1 and Ni-CAT-1, but they fail to synthesis Cu-CAT-1 thin film of the same quality. Nevertheless, in this regard, the coordination of $Cu^{2+}$ ions with HHTP ligand for the thinnest (10 nm thick) 2D Cu-CAT-1 MOF film sequential deposition of nanosheets onto the surface was already reported for the enhanced electrical response by Carlos and co-workers [99]. **Figure 21** Schematically demonstrated the surface anchored deposition of nanosheets. After the formation of the self-assembled monolayer, a well-oriented, conductive, ultrathin films of Cu-CAT-1 were obtained using bottom-up techniques.

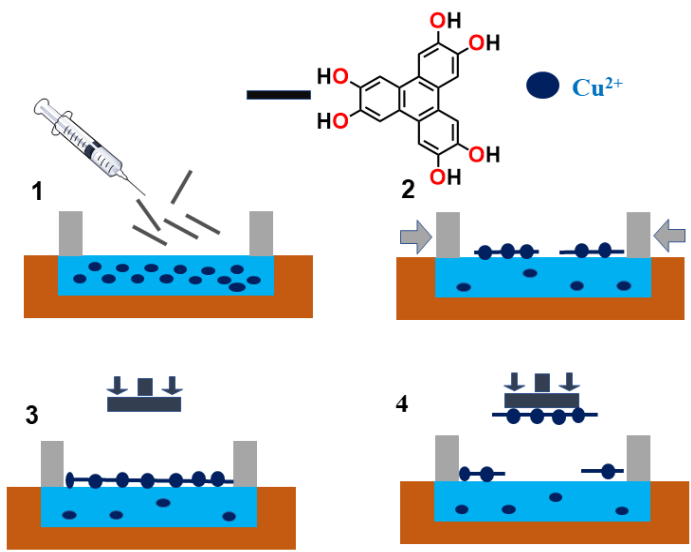

**Figure 21.** Synthesis of ultrathin homogenous films by using four-step method (1) formation, (2) compression, (3) optimal pressure for the continuous layer, (4) finally transfer to the substrate using Langmuir-Schaefer method.

For charge transport studies, a thin film-based device was fabricated with bottom-contact geometry, using pre-patterned Au electrodes (500 μm width) as the bottom contact. On top $SiO_2$ substrate (230 nm thick) and highly doped n-type Si was used as a dielectric and gate, respectively **(Figure 22a).** Optical microscopy confirms the full device coverage after transfer of Cu-CAT film, as shown in (**Figure 21b**). In **Figure 21c**, the linear increment in the conductivity at room temperature revealed ohmic contact between the thin film and gold electrode. In addition, temperature-dependent studies **(Figure 21d)** distinguished that cooling (above 240 K) progressively decreased the conductivity and furnished the thin film's semiconducting nature. Upon further cooling (below 240K), a non-linear dependence of the conductivity on temperature was reached.

The 2D conducting MOF $Ni_3(HITP)_2$, which consists of 2,3,6,7,10,11-hexaaminotriphenylene linkers and nickel metal ions, shows high electrical conductivity (>5000 $Sm^{-1}$), has been well-explored in the supercapacitors, Battery etc.[100,101]. For its more widespread use in device fabrication, a thin film of good orientation, homogeneous, controlled thickness, minimal roughness, and defect-free is extremely desirable. Towards that goal, recently Ohata et. al synthesized well oriented thin film of $Ni_3(HITP)$ MOF by employing an excellent air/liquid (A/L) bottom-up approach **(Figure 22)**[102]. The authors also contend the first report of the synthesis of the unidirectional orientation, crystalline thin film ensured by synchrotron X-ray crystallography. The thin film can be easily transferred onto any substrate which makes it superior to the solvothermal or vapor-induced conversion method in which the it can be grown only on a particular substrate. The planar electrical conductivity was measured by inserting the thin film between gold electrodes **(Figure 23a)**. Interestingly, the electrical conductivity was found to be 0.6 S $cm^{-1}$ for the 70 nm thickness, which was the highest among the previously reported nanosheets having a thickness of less than 100 nm. Correspondingly, in the I-V plot, the linear variation of current with voltage demonstrates the metallic property of the thin film **(Figure 23 b)**. These mesmerizing features were attributed to the uniaxially-oriented thin films which were achieved by the novel A/L interface technique. This discovery paves the way for the design of a highly efficient electronic device by controlling such highly oriented nanosheets.

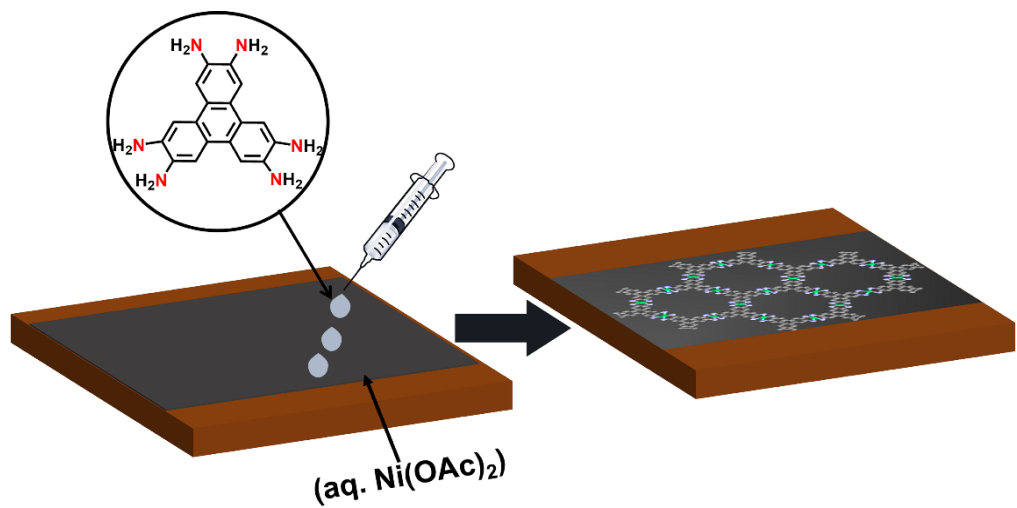

**Figure 22**. Synthesis of a thin film of HiTP-Ni by employing the air/liquid interface technique. A solution containing 2,3,6,7,10,11-hexaaminotriphenylene (HATP) in methanol is uniformly distributed over an aqueous solution of Ni(OAc)₂ 4H₂O in the Langmuir trough. Subsequently, the reaction of HATP and Ni(II) ions at the air/liquid interface leads to the formation of a HiTP-Ni thin film. This synthesized thin film can be easily transfer to any substrate (e.g. Si/SiO₂, gold , silver, ITO, FTO) according to our need.

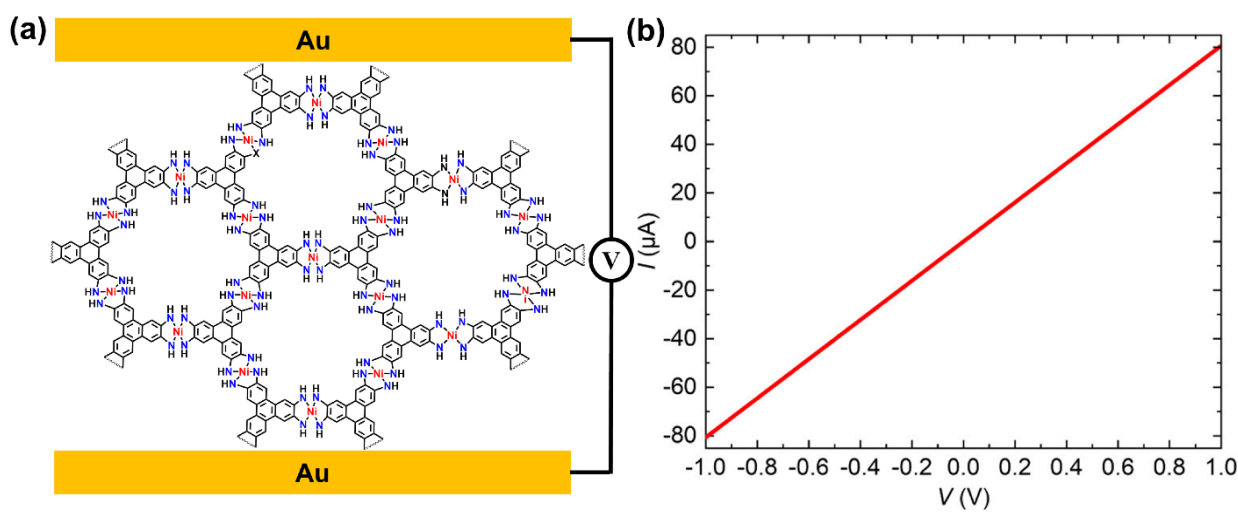

**Figure 23**. (a) Schematic illustration of device Au/HiTP/Au (b) The linear variation of current with respect to voltage in the I-V plot disclose the metallic character of the synthesized thin film. Reproduced with permission from ref.[102], copyright 2021, American Chemical Society.

However, 2D-MOFs have been used successfully in fabricating real-world applications. Nevertheless, its charge transport mechanism is not well understood, as mostly polycrystalline or thin films are investigated, and studies based on single crystals. Recently Day *et al.* extensively explored the electrical property of the suspension of Ni₃(HITP)₂ onto a silicon substrate, and the top electrode (Pd/Ti) was deposited by employing electron-beam lithography [4]. **Figure 25a** and **b** show the schematic diagram of the 2D MOF device and the corresponding SEM images, respectively. The electrical conductance value was obtained at 1.3 ×10⁻³ S at 295 K and 0.7 ×10⁻³ S at 1.4K (Single rod, Four Probes) and 0.25

×10$^{-3}$ S at 295 K and 0.13 ×10$^{-3}$ S at 1.4K (Single rod, two probe) as displayed in **Figure 25c**. On further reducing the temperature to 0.3K slight decrease in conductance was observed. That implies that MOFs at extremely low temperatures (~0.3 K) function as a metal, whereas above that, semiconductors. Moreover, the electrical conductivity of a single crystal was found higher than that of its polycrystalline thin film, owing to the lack of grain boundary contribution. This study has given insight into charge transport.

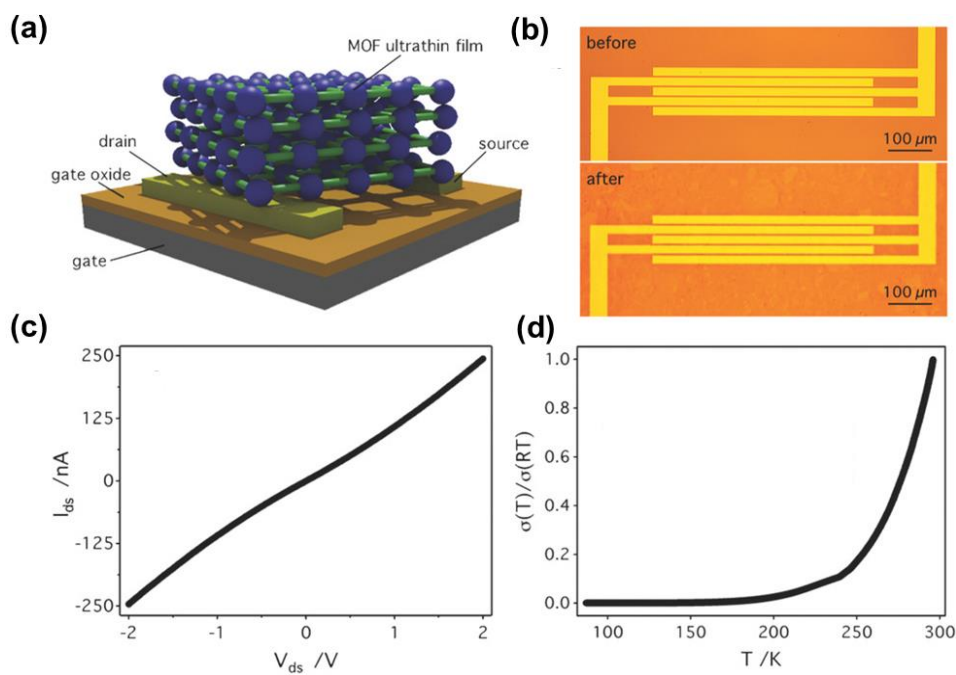

**Figure 24.** (a) Schematic illustration of Cu-CAT-1thin films (10 nm) device having bottom-contact geometry. (b) Optical microscope image of the real device before and after loading thin films. (c-d) I-V plot measured at room temperature and under different temperatures, respectively. Reproduced with permission from ref. [99], copyright 2018, Wiley-VCH.

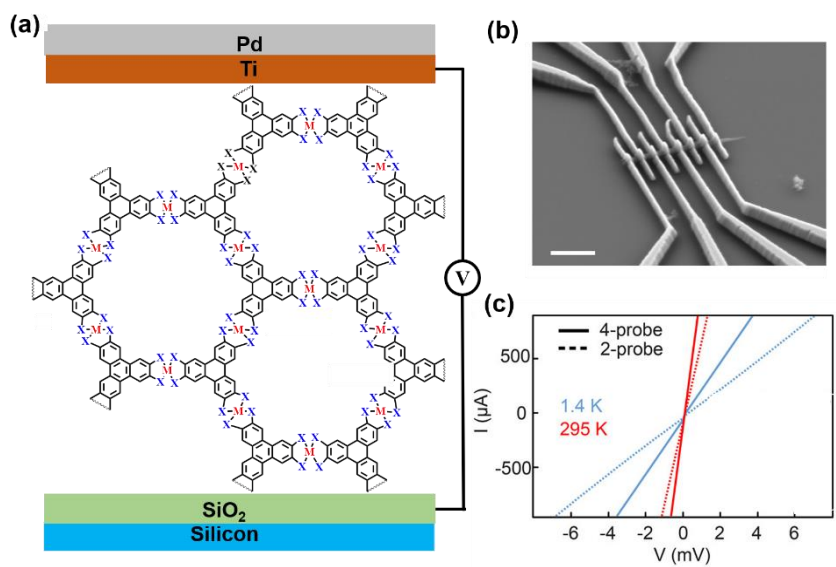

**Figure 25.** (a) schematic illustration of molecular junction Pd/Ti/Ni$_3$(HITP)$_2$/SiO$_2$/Si. (b) Scanning electron microscope image of the real device. (c) I-V plot at 1.4 K and 295 K using 2-probe and 4-probe. Reproduced with permission from ref. [4], copyright 2019, American Chemical Society.

Recent work by Dincă and co-workers discussed the structure-function relationship of HHTP ligands with lanthanide cations due to its empty $5d$ shells as well as the reluctance of 4f orbitals participating in bonding, which makes $Ln^{3+}$ ions form noticeably more ionic compounds than other $1^{st}$-row transition metals [103]. In that work, they have correlated the stacking distance, as a function of the size of the lanthanide cation with that of the two-probe conductivity and the optical bandgap of MOFs. They synthesized a series of lanthanide-based MOFs, LnHHTP (Ln =Yb, Ho, Nd and La) followed by charge transport studies. Surprisingly, **Figure 26a** revealed that conductivity was dominated by stacking distance rather than the extent of covalency between metal ion and ligand. YbHHTP and HoHHTP, which consist of comparatively small metal ions and have more compact stacking, exhibited conductivity (~ 0.05 Scm⁻¹) almost three-fold higher than that of LaHHTP and NdHHTP (~$0.9\times10^{-4}$ S cm⁻¹). This result suggests charge transport is taking place out of the plane instead of in-plane. Additionally, through temperature-dependent studies, similar activation energy was found for all MOFs.

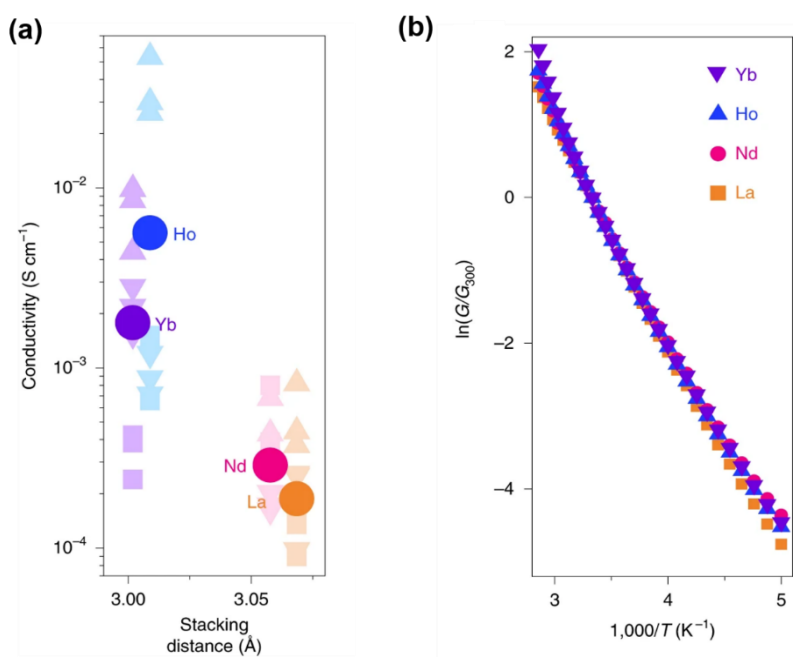

**Figure 26.** (a) Two-probe electrical conductivity of LnHHTP (Ln =Yb, Ho, Nd and La) with respect to stacking distances at 29 °C. Squares, downward-facing triangles, and upward-facing triangles represent three independent batches of each sample of distinct pellet measurement. (b) Temperature dependence the normalized conductance of four different LnHHTP samples based on Arrhenius coordinates, showing similar temperature-activated conductivity. Reproduced with permission from ref. [103], copyright 2020, Springer Nature.

Besides, to unravel the temperature-dependent electrical conductivity Dincă and coworkers synthesized three conductive MOFs using mixed-valent tetrathiafulvalene tetrabenzoate (TTFTB) ligand and $La^{3+}$ metal ions through the alteration of temperature and solvent ratio via the solvothermal method (**Figure 27a**) [104]. The comparative charge transport studies were performed by sandwiching pellets between carbon paste and gold wire (**Figure 27b**). The rate of charge transport was found directly proportional to the extent of π-π overlapping and inversely proportional to S⋯S distance rendering hopping mechanism. **Figure 27c** shows the device configuration. Interestingly, lanthanide-based MOFs have been exploited as an excellent tunable emitting material through doping the framework with other elements. Zhu *et al.* developed new double-chain-based 3D lanthanide MOFs exhibiting proton conduction and tunable emission property. The MOFs manifest a proton conductivity of $1.6 \times 10^{-5}$ S cm$^{-1}$ at 75 °C and tunable white light emission [105]. In recent review, Han and co-workers summarised the role of lanthanide MOFs and their derivative as catalysis, including heterogeneous organic catalysis and photocatalysis [106]. However, the purpose of lanthanide in MOF towards thin-film device integration has not yet been debated.

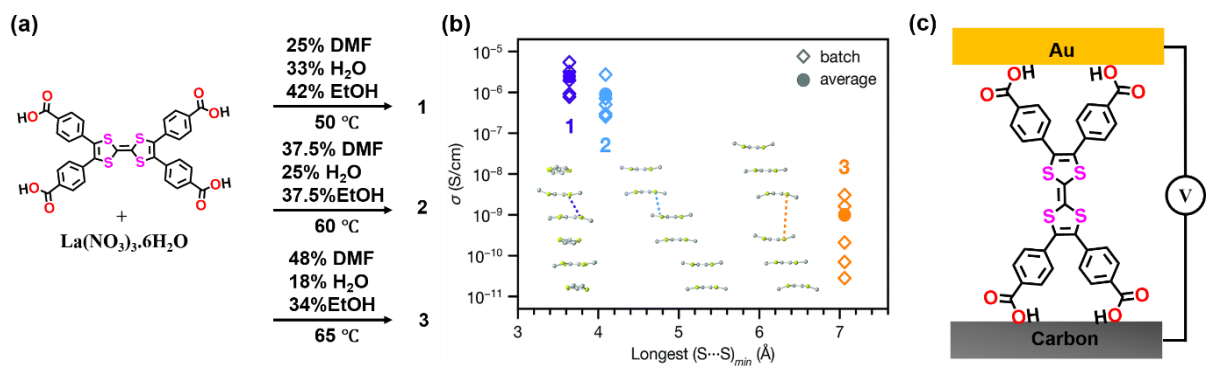

**Figure 27.** (a) Scheme for synthesizing three electrically conductive MOFs (1, 2, 3) by variation of temperature and solvent ratio. (b) Conductivity *vs.* Longest (S⋯S)$_{min}$ plot of three MOFs. (c) Schematic illustration of device Au/H$_4$TTFB/Carbon molecular junctions. Reproduced with permission from ref. [104], copyright 2019, The Royal Society of Chemistry.

### 7.2 Charge Transport studies in the 3D MOF

The synthesis of 3D electrically conductive MOFs is a big challenge due to their high porosity. Therefore, to date, only a small number of such MOFs have been reported. Most recently, Medina and co-workers reported a microporous 3D electrically conducting MOF consisting of HHTP linker and Fe (iii) metal node having porosity 1400m$^2$/g (**Figure 28**) [107]. The conductivity value was obtained for Fe-HHTP-MOF, about $10^{-3}$ Scm$^{-1}$, which is higher than that observed for any other 3D MOFs so far. The quantum mechanically charges transport studies revealed electron transfer taking place through Fe$^{II/III}$ and semiquinone/catechol redox center.

Skorupsii *et al.* recently synthesized intrinsically electrically conductive and structurally isotropic MOFs, Eu$_6$HOTP$_2$ and Y$_6$HOTP$_2$ [108]. These MOFs consist of 2,3,6,7,10,11-hexaoxytriphenylene (HOTP) linker and Eu and Y rare earth metal nodes, respectively. These MOFs were investigated using a two-probe method, subsequently activating by heating at 90°C for 18 hours under dynamic vacuum conditions. The MOFs showed electrical conductivity in the order of $10^{-6}$ to $10^{-5}$ S/cm and surface area 780 m$^2$/g. Activation of MOFs led to partial oxidation of catechol moieties of HOTP$^{6-}$, yielding charge carrier semiquinone radical as a free charge carrier. **Figure 29** shows the diffuse reflectance spectra to establish the heating effect under inert conditions and the corresponding I-V plot.

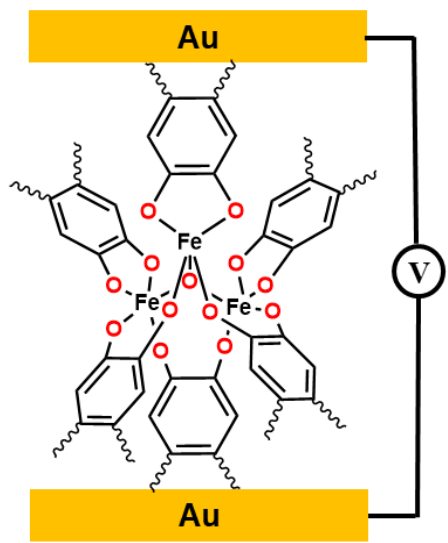

**Figure 28.** Schematic illustration of molecular junction Au/Fe-HHTP/Au. The junction was tested for the charge transport studies using the two-probe as well as four-probe method.

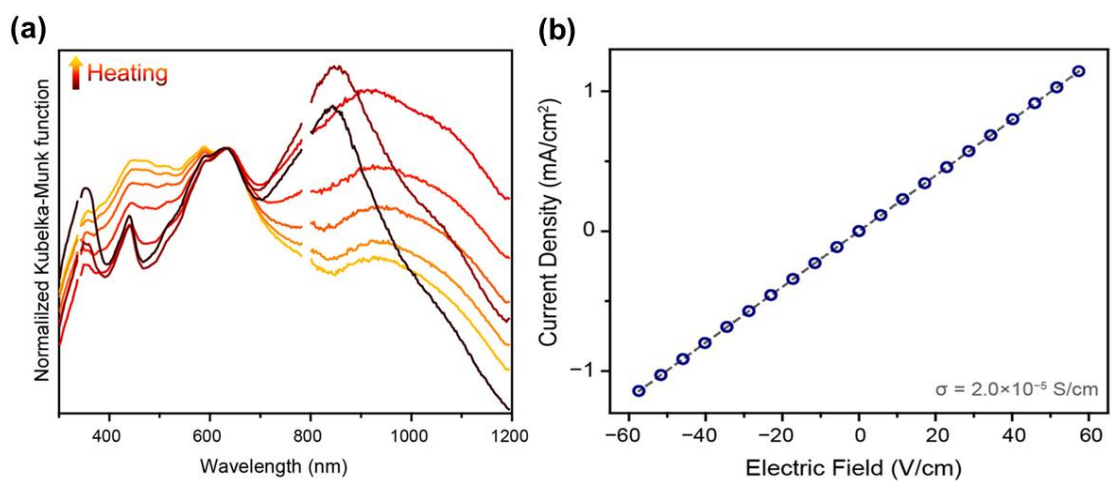

**Figure 29.** (a) Diffuse reflectance spectra of $Y_6HOTP_2$, obtained from heating the sample under the inert condition of the dinitrogen atmosphere (black, initial spectrum; yellow, final spectrum). (b) I-V plot of $Y_6HOTP_2$ measured electrical conductivity was $2.0 \times 10^{-5}$ S/cm using the two-probe method. by Diffuse reflectance spectra, the presence of new broadband at 1000-1200 nm advocates the presence of semiquinone radicals. Reproduced with permission from ref. [108], copyright 2020, American Chemical Society.

Sengupta *et al.* made 3D and 2D Cu-porphyrin ($H_2TPyP$) MOFs by altering the solvent (like tetrachloroethane for 3D and chloroform for 2D MOFs, respectively) with the moiety and $Cu^{2+}$ as the metal ion linker [109]. $H_2TPyP$ ligand is better known for employing the specific host-guest chemistry due to its electron-accepting property (p-type) and high surface area [110]. To check the electrical conductivity, they converted the 2D MOF into film and doped it with TCNQ. The doped thin film exhibited 3 orders of magnitude higher conductivity ($1 \times 10^{-4}$ $Sm^{-1}$) concerning the undoped one ($10^{-7}$ $Sm^{-1}$). Additionally, the dinuclear Cu(II) undergoes magnetic superexchange interactions between copper atoms into the bridging organic linker. The TCNQ-doped 2D MOF showed a higher magnetic moment value due to the appearance of the carbon radical in the charge-transfer complex built between TCNQ and 2D MOF.

## 8. Various examples of MOFs-based electronics devices

Research on MOFs-based molecular electronics has been drastically advanced in recent years ranging from diodes, photoelectric effect, memory devices, field-effect transistor (FET), rectifier, etc [111,112]. Recent advancements made with the MOFs materials in designing memristive devices. MOFs have been utilized in the circuit as active electronic elements (AEEs) to operate the memristive devices at the molecular level. Working principles of memristive or memory-based resistive devices are entirely different from the CMOS-based technological devices, as the formers work on interfacial charge transfer and metallic ion transport, while the latter relies on electron transport. The memristive devices follow SET/RESET (S-R) operations triggered by electric field or magnetic field by the principle of variable mobility of the ions involved in the said devices [113]. Thanks to the electrical insulation of the MOF, Yoon *et al.* reported the first example of an external electric field-driven γ-cyclodextrin based Rb-CD-MOF for resistive memory device application owing to the nanoscopic porous network for facile transportation of ions/charges [114]. Recently, Pham and co-workers studied volatile threshold as well as non-volatile memory devices based on zirconium (IV)-carboxylate-based large surface area MOF (UiO-66). The device (Ag/MOF with PVA conjugation/FTO) exhibits multi-resistive switching behavior at the relatively low operating voltage and the bipolar resistance state at the high working voltage, having high ON/OFF ratio ($\sim 10^4$), excellent endurance ($5 \times 10^2$ cycles), and longer retention time ($\sim 10^4$ s) which considered as the excellent characteristics of a memory device [115]. The FET is an applied electronic research domain where, electric field-driven semiconductor device, consists of three terminals, i.e., *gate, source, and drain*. It has played an enormous role in exploring charge transport properties (charge carrier type and mobility value) and the fabrication of several electronic devices. Benefiting from high crystallinity, defect-free thin film, and conductivity in the semiconductor range, MOFs have emerged as an appealing candidate for FET fabrication. In this context, Wang *et al.* prepared $Ni_3(HITP)_2$ membrane as the channel material of the FET growing in $Si/SiO_2$ wafer with pre-patterned Au electrodes through the in situ formed solid-liquid interface[116]. Owing to the excellent coordination between the HITP ligand and Ni metal center and the π-π interaction among the stacked layers the Ni-MOF-FET shows high carrier mobility of 45.4 $cm^2 V^{-1} s^{-1}$ and $I_{on}/I_{off}$ ratio of $2.29 \times 10^3$. Additionally, the material operated as a liquid-gated device with bipolar behavior and reasonable response as a biosensor for the gluconic acid detection within the range from $10^{-6}$ to $10^{-3}$ g/mL. Another intriguing utilization of MOF as a rectifier was executed by Ballav and his research team [117]. They prepared a thin film (thickness ~700 nm) of Cu (II)-1,2,4,5-Benzene Tetracarboxylic acid (Cu-BTEC) on the gold substrate by employing the layer-by-layer (LBL) method. On doping the thin film with TCNQ, the only top layer was doped, whereas the bottom layer closer to Au substrate persisted undoped due to saturation of TCNQ uptake. As a result, while carrying out I-V measurement in-plane only, improvement of the electrical conductivity was simple. In contrast, during cross-plane measurement, the conductivity enhancement was not the simple as earlier. Since the bottom part is not a good conductor, the current flow was controlled due to electronic heterostructure, resembling a typical p-n junction with a rectification ratio $\geq 10^5$ that is compatible with an inorganic-based semiconductor (**Figure 30**). In another work, Ballav and Co-workers reported the synthesis of Cu-TCNQ based SURMOFs which accomplished the prerequisite of device fabrication having contact angle ~140⁰ (reveals hydrophobic nature) [118]. Interestingly, exposure in $I_2$ vapor shows a notable rectification factor of ~$10^2$ and additionally showed non-ohmic conductivity in the range of $10^{-5}$ S $cm^{-1}$. The leading cause behind the electrical rectification in the Cu-TCNQ system is the ligand-centered redox reaction caused by $I_2$. So far, the methods we have come across also can be an additional platform for designing spintronics devices for nanotechnology.

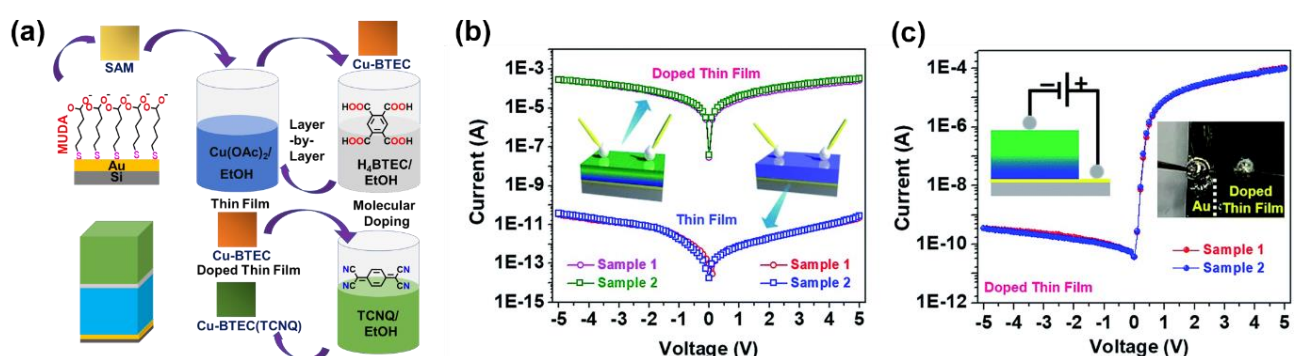

**Figure 30.** (a) Schematic view of the fabrication steps of Cu–BTEC thin film using the LBL method. After ten repeating cycles, thin-film was doped with TCNQ. (b) I-V plots (in-plane measurement) of a thin film of Cu-BTEC (blue and red) and doped (green and magenta). A considerable enhancement in the electrical conductivity was obtained upon doping. (C) Illustrating I-V plots (blue and red) of doped thin film (cross-plane measurement), having rectification with an RR value $\geq 10^5$; insets image corresponds to the scratched sample till Au has been visible for the I-V measurements. Reproduced with permission from ref. [117], copyright 2018, The Royal Society of Chemistry.

Recently, Halder *et al.* showcase the optoelectronic properties of MOFs, which are rarely explored applications of conducting MOFs, mainly using non-transition elements such as Cd ($d^{10}$ electronic configuration) [112]. Thanks to zero crystal field stabilization energy (CFSE), $d^{10}$ system can adopt any coordination geometry. The coordination number may differ from four to eight. Hencethey synthesized Cd(II) based 2D MOF, [Cd(4-bpd)(SCN)$_2$]$_n$ where 4-bpd = 1,4-bis(4-pyridyl)-2,3-diaza-1,3-butadiene as a photosensitive Schottky diode. The two probe-based DC conductivity studies revealed a non-linear rectifying function on both under dark and incident light irradiation, which imports the Schottky diode nature. The device has been made of ITO/[Cd(4-bpd)(SCN)$_2$]$_n$/Al with a total thickness of the molecule 1 μm (**Figure 31**). They found high electrical conductivity around $2.90 \times 10^{-4}$ $Sm^{-1}$ and $7.16 \times 10^{-4}$ $Sm^{-1}$ for both dark and illumination conditions, respectively. Theoretically, they estimated less energy is needed for the electronic transition between the valence band and conduction band upon illumination of incident light. The irradiation causes little change in the bond distance between nitrogen-nitrogen, nitrogen-carbon, and carbon-carbon of the two adjacent pyridyl groups of the 4-bpd ligand.

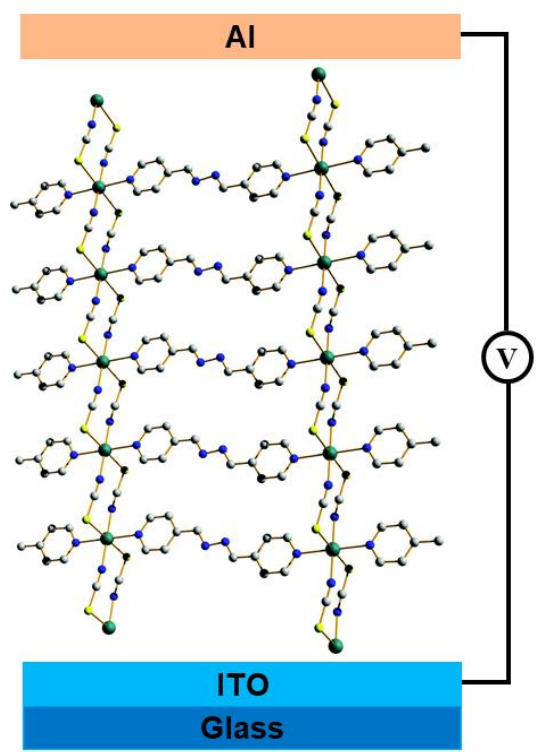

**Figure 31.** Schematic diagram of molecular junction ITO/[Cd(4-bpd)(SCN)$_2$]$_n$/Al. The junction has been tested for the dc conductivity measurement using the two-probe method. The molecule inside the junction represents ladder-type geometry.

## 9. Conclusions and future perspective

In summary, recent advances made with diverse MOF thin films concerning developing synthesis strategies and electrical property benefit electronic device applications. The traditional reticular chemistry of MOFs has depended on the construction of communicated inorganic strong metal-ligand bonds to get the desired structure and properties. Last five years, the surface-related chemical processes of thin-film MOF has increased expressively, facilitating a range of applications. However, despite the plethora of success, control over the morphology, crystallinity, roughness, adjustable conductivity, and orientation of the films into electronic devices remains challenging. Considering this review, we have mostly emphasized the π-π interactions between the ligands and the corresponding interlayer charge transport property to unravel the structure-function relationship. Solvent modulated reversible electrical conductivity variation by coordinating the solvent to metal sites also established the structure-function relationship through the changes in chemical composition. The charge transport mechanism relied on the hopping regime, was also highlighted to find the high-electron conduction pathway by selecting the example of intercalated MOF materials. For this purpose, the external electric field-dependent reversible manipulation of long-range order and 3D porous network in MOFs is imperative and needs extensive study. The electric field as a stimulus is significant since it offers a manipulative range of field strength, and their localization is relatively facile, which can overcome interference. The control over MOF particle orientation in a reversible manner under an external electric field is even more challenging while retaining most of the bulk properties alongside. Integration of MOFs onto suitable substrates for electronics and electro-optical devices and their electric field manipulation have seen only a little success. We anticipate that upon improving the physical, chemical, optical, electrical behavior of MOFs and enhanced engineering, the devices for real-world applications in electrical switching, flexible electronics, and photonics optoelectronics are not far away. However, there are many voids in the MOFs-electronics that need further studies to enhance understanding of the actual landscape of charge-transport phenomena, which is urgent for taking them out of the four-wall lab. Space to the household. While conjugated conducting polymers and some organic semiconductors show superior charge transport properties and relatively higher carrier mobility of ~1-10 $cm^2$ $V^{-1}$ $s^{-1}$, which is much higher than amorphous silicon [119]. However, more underlying solid-state devices physics must be understood in enhancing the properties in MOFs-based materials. MOFs are composed of many atoms arranged periodically leading the formation of intriguing complex structures that impart many physical, chemical, electronics, optoelectronics that can be combined for technological applications.

i. ***Electrical Conductivity of Heterostructures MOFs:*** Heterostructures MOFs comprised of either two different MOFs or MOFs with different metal ions would be highly important in their electrical properties. They can behave as electrical diodes provided different frontier molecular orbitals (HOMOs and LUMOs) of the MOFs materials. Their electron wave function study by quantum mechanical method would also be exciting as theoretical support may find their (de)localization of the electrons in two MOFs interfaces also at the electrode/MOFs$_1$/MOFs$_2$/electrode interfaces.

ii. ***Role of electrodes in devices fabrication:*** It is indisputable that electrodes or conductors play crucial roles in controlling the charge-transport properties of molecular electronic devices. Since different electrode has different Fermi level, thus the energy barrier height which is the difference of $E_{HOMO/LUMO}$-$E_F$ controls the charge injection step and charge detection to the other end of the electrode [120]. Molecular rectification can be observed with the electrodes of varied work functions because $E_{HOMO/LUMO}$-$E_F$ can be lower at one electrode, while can be higher at the second electrode. The study must be made with symmetric electrodes followed by heteroelectrodes by altering bottom and top contact.

iii. ***Impedance spectroscopy studies on MOFs devices:*** Recent advancements made with (semi)conducting MOFs are fascinating; however, no straightforward efforts have been emphasized on studying individual junction's electrical components such as contact resistance, charge-transfer resistance, molecular junction resistance, and capacitance. These circuit elements are crucial in MOFs-based electronics devising to easily understand how these components affect the charge-transport phenomena. Besides, there is no modeling based on either experimental and/or theoretical data. These data would expedite device modeling such as Randles equivalent circuit modeling [121].

iv. ***Thickness-dependent MOFs devices***: To our surprise, charge-transport phenomena in varied thicknesses of the same MOFs have not been executed so far. However, these types of studies are necessary to examine how thickness can affect the charge-transport properties that can be measured under identical conditions i.e. fabrication conditions, bottom and top electrode, the measurement set up must be kept unchanged [74]. Likewise, the same SURMOFs of different thicknesses must be employed in device stacking for charge-transport studies under temperature, the electric field that would for sure felicitate.

v. ***Metal-free electrodes in device stacking***: The metal-free electrode is highly desirable for real-world application as metals are highly costly, need metal depositors to prepare metallic electrodes. In this scenario, conducting carbon can be an exciting alternative to replace metal electrodes [122–125]. A hot off the press review depicting recent progress on carbon-based molecular electronics, where carbon can be employed as both 'bottom' and 'top' contact has been recently published by our research group [37]. Thus, the devices stacking can be changed from metal/MOFs/metal (M/M/M) to Carbon/MOFs/Carbon (CMC) configuration. Carbon has a Fermi level alike many metals, for instance, Au. It is cost-effective, patternable, and deposited even on flexible substrates. It can impede filament formation in the course of top contact deposition. The transparency makes it a better candidate for optoelectronic studies on the same platform. Other possibilities such as ITO, PET, FTO can also be used to replace metal contacts.

vii. ***Stimuli-responsive electrical conductivity in MOFs-based devices***: A polar solvent can enhance electrical conductivity, as a solvent-mediated interface creates an asymmetric internal electric field. If we find a way to make them reversible conduction changes, that would be fantastic. Light can trigger electrical conductivity; hence, the MOFs community needs to design such photoactive MOFs that can reversibly (probably by UV and Visible light) interplay with light causing ON/OFF conducting switching, which can be a consequence of MOFs-based diode.

viii. ***Electrochemical studies using MOFs***: Adequate electrochemical measurements are necessary to know the redox properties of MOFs. The reliable electrochemical analysis would facilitate in understanding their double-layer capacitance, charge-transfer rate constant, data To date, few redox-active MOFs are known [126–128]. Recent theoretical and experimental work reveals that MOFs can be good electrode materials [129]. Yet, the underpinning mechanism is needed.

viii. ***MOFs-based spintronics***: MOFs can also be embedded on ferromagnetic electrodes followed by the same or different ferromagnetic electrode deposition as the top contact. MOFs can be deposited on ferromagnetic substrates in enormous ways [31,130,131]. In recent years, significant efforts have been devoted to synthesizing chiral MOFs and it would be exciting to examine whether they act as spin filtering [132–134].

ix. ***Theoretical Frameworks are looked for***: Many theoretical models have been made with organic crystalline semiconductors for their charge-transport and high mobility phenomena [135–139]. In the case of the band pictures of organic materials, the mean free path of the charge carriers' collisions is predicted significantly longer than the spacing among the corresponding lattice sites, thus a higher diffusion length is accessible. According to the theoretical frameworks, the electronic transfer rate dominates over the scattering rate, still, the experimental charge-carrier mobility in organic semiconductors is much less than the theoretically predicted as high as 50 cm$^2$ V$^{-1}$ s$^{-1}$ considering intermolecular spacing of 6Å [140,141]. The recent progress made with MOFs enhanced electricidal conductivity is quite fascinating. Although, their origin of such monumental electrically conductive is still unexplored. In a recent review, Maurin and a co-worker discussed structural and electronic responses in Zeolitic imidazolate frameworks (ZIFs) [142]. They have emphasized the state-of-the-art experimental and computational procedure accomplishing the stimuli-responsive behavior such as temperature, mechanical pressure, restoration, or electromagnetic fields of four prototypical ZIFs like ZIF-8, ZIF-7, ZIF-4, and ZIF-zni. Thus, the theoretical framework can trigger a deep understanding of the underlying physical, chemical, and electronics properties governing overall conductivity enhancement in the MOFs. Identification of steppingstones with experimental evidence would be instrumental in understanding the charge propagation in MOFs. So, the detailed narration of electron propagation in MOFs can encourage the experimental MOFs community to resolve those drawbacks that impede the conductivity.

Despite many exciting research works, we still believe monumental, and landmark experiments are needed with varied MOFs to commercialize them as electronic gadgets. The MOFs community should pay sincere attention to eliminating interfaces and structural defects inhibiting charge flowing coherently. Many research groups have been well-known for creating conductive MOFs and devised varied experimental techniques to probe the conductivity of MOFs using two probes, four-probe methods. Thus, the MOFs community needs real blueprints for electronics applications. This review can motivate the MOFs community to systematically discover the charge-transport phenomena in diverse chemical, structural MOFs.

## List of Abbreviations

| | |
|---|---|
| A | Junctions area (in cm$^2$) |
| AC | Alternating current |
| AEEs | Active electronic elements |
| ALD | Atomic layer deposition |
| BDC | 1,4-benzene dicarboxylate |
| CAFM | Conducting atomic force microscopy |
| CAMs | Covalent-assembled monolayers |
| CB | Conduction band |
| Cdl | Double layer capacitance ($F/cm^2$) |
| CFSE | Crystal Field Stabilization Energy |
| C/M/C | Carbon/molecules/carbon |
| CMOS | Complementary metal-oxide–semiconductor |
| CP | Conjugated polymers |
| CPs | Coordination polymers |
| CV | Cyclic voltammetry |
| CVD | Chemical vapor deposition |
| DC | Direct current |
| DFT | Density function theory |
| DMF | Dimethyl formamide |
| DOS | Density of states |
| DSBDC | 2,5-disulfidobenzene-1,4-dicarboxylate |
| d | Thickness of the molecular layers between two conductors |
| ECD | Eltrochromic devices |
| ECMOFs | electrically conductive metal-organic frameworks |
| E-field | Electric field (V/cm) |
| E/M/E | electrode/molecules/electrode |
| EIS | Electrical impedance spectroscopy |
| ET | Electron transfer |
| $E_a$ | Activation energy (eV) |
| $E_F$ | Fermi levels (eV) |
| $E_g$ | Band gap energy (eV) |
| E-grafting | Electrochemically grafting |
| Fc | Ferrocene [Fe(cyclopentadiene)$_2$] |
| FM | Ferromagnet |
| FMOs | Frontier molecular orbitals |
| FTO | Fluorine doped tin oxide |
| FESEM | Field-emission electron microscopy |
| HHTP | 2,3,6,7,10,11-hexahydroxytriphenylene |
| HKUST | Hong Kong University of Science and Technology |
| HOMO | Highest occupied molecular orbitals |
| iMOFs | Intercalated metal-organic frameworks |
| ITO | Indium tin oxide |
| IoT | Internet-of-thigs |
| J | Current density ($Amp/cm^2$) |
| $J_o$ | pre-factor |
| $K_B$ | Boltzmann constant |
| $k_{et}$ | Electron transfer rate |
| LbL | Layer-by-layer |
| LPE | Liquid-phase epitaxy |
| LUMO | Lowest unoccupied molecular orbitals |
| M-CAT-1 | metal-catecholate |
| MD | Molecular dynamics |
| MEDs | Molecular electronics devices |
| M/I/M | Metal/insulator/metal |
| MJs | Molecular junctions |
| MMLs | Molecular multilayers |
| M/M/M | Metal/molecules/metal |
| MLs | Molecular layers (nm) |
| MOFs | Metal-organic frameworks |

| PET | Polyethylene terephthalate |
| PPF | Pyrolyzed photoresist film |
| PVD | Physical vapor deposition |
| PEDOT | Poly(3,4-ethylenedioxythiophene) |
| R | Resistance (in Ohm) |
| RCA | Radio Corporation of America |
| $R_{ct}$ | Charge-transfer (or molecular) resistance (in Ohm) |
| $R_s$ | Contact resistance (in Ohm) |
| S | Siemence ($Ohm^{-1}$) |
| SAMs | Self-assembled monolayers |
| SCPs | Surface coordination polymers |
| SEM | Scanning electron microscopy |
| SURMOFs | Surface-confined metal–organic frameworks |
| T | Temperature (in K) |
| TCNQ | 7,7,8,8-Tetracyanoquinodimethane |
| TDDFT | Time-dependent density functional theory |
| THT | 2,3,6,7,10,11-tripheylenehexathiolate |
| TLs | Template layer |
| TTFTB | tetrathiafulvalene tetrabenzoate |
| UPS | Ultraviolet photoelectron spectroscopy |
| VAC | vapor-assisted conversion |
| VB | Valence band |
| vdP | Van der Pauw |
| WF | Work function (eV) |
| XPS | X-ray photoelectron spectroscopy |
| ZIF | Zeolitic imidazolate framework |
| $\alpha$ | Potential well height (eV) |
| $\beta$ | Charge-transport exponential decay constant |
| $\sigma$ | Electrical conductivity ($S\ cm^{-1}$) |
| $\Gamma$ | Surface coverage ($moles/cm^2$) |
| $\phi_T$ | Energy barrier for coherent tunneling (eV) |

**Declaration of competing interest:** The authors declare no competing interest in this review article.

## Acknowledgments

PJ acknowledges IIT Kanpur for providing Institute postdoctoral fellowship. RKP gratefully acknowledge the junior research fellowships from the Council of Scientific and Industrial Research (CSIR), New Delhi. PCM acknowledges a start-up research grant support from the Department of Science and Technology, New Delhi, India (SRG/2019/000391) and an initiation research grant for independent laboratory setup from IIT Kanpur (IITK/CHM/2019044).